\documentclass[12pt]{article}

\usepackage{natbib}
\usepackage{bm}
\usepackage{amsmath}
\usepackage{amsfonts}
\usepackage{url}
\usepackage{floatflt}
\usepackage{graphicx}
\usepackage{color}

\newcommand{\lesssim}{\mathrel{\hbox{\rlap{\hbox{\lower4pt\hbox{$\sim$}}}\hbox{$<$}}}}
\newcommand{\gtrsim}{\mathrel{\hbox{\rlap{\hbox{\lower4pt\hbox{$\sim$}}}\hbox{$>$}}}}
\newcommand{\uvec}[1]{\widehat{\rm #1}}
\newcommand{\radon}[1]{\widehat{#1}}

\begin{document}

\title{Non-Baryonic Dark Matter}
\author{Paolo Gondolo\\ \it\small Department of Physics, University of Utah, \\ \it\small 115 South 1400 East, Suite 201, Salt Lake City, Utah 84112, USA}
\date{~}
\maketitle

These lectures on non-baryonic dark matter matter are divided into two parts. In the first part, I discuss the need for non-baryonic dark matter in light of recent results in cosmology, and I present some of the most popular candidates for non-baryonic dark matter. These include neutrinos, axions, neutralinos, WIMPZILLAs, etc.
In the second part, I overview several observational techniques that can be employed to search for WIMPs (weakly interacting massive particles) as non-baryonic dark matter. Among these techniques, I discuss the direct detection of WIMP dark matter, and its indirect detection through high-energy neutrinos, gamma-rays, positrons, etc.
 References cited in
these lectures are intended mostly for further study, and no attempt has been made to provide a comprehensive list of original and recent work on the subject.

\section{The need for non-baryonic dark matter}

We live in a time of great observational advances in cosmology, which have given  us a consistent picture of the matter and energy content of our Universe. Here matter and energy (which special relativity tells us are equivalent) are distinguished by their different dependence on the cosmic volume: matter density decreases with the inverse of the volume, while energy density remains (approximately) constant.

Nothing is known about the nature of the energy component, which goes under the name of dark energy. Of the matter component, less than 2\% is luminous, and no more than 20\% is made of ordinary matter like protons, neutrons, and electrons. The rest of the matter component, more than 80\% of the matter, is of an unknown form which we call non-baryonic. Finding the nature of non-baryonic matter is referred to as the non-baryonic dark matter problem. 

\begin{figure}[t]
\label{fig:1}
\centering
\includegraphics[width=0.8\textwidth]{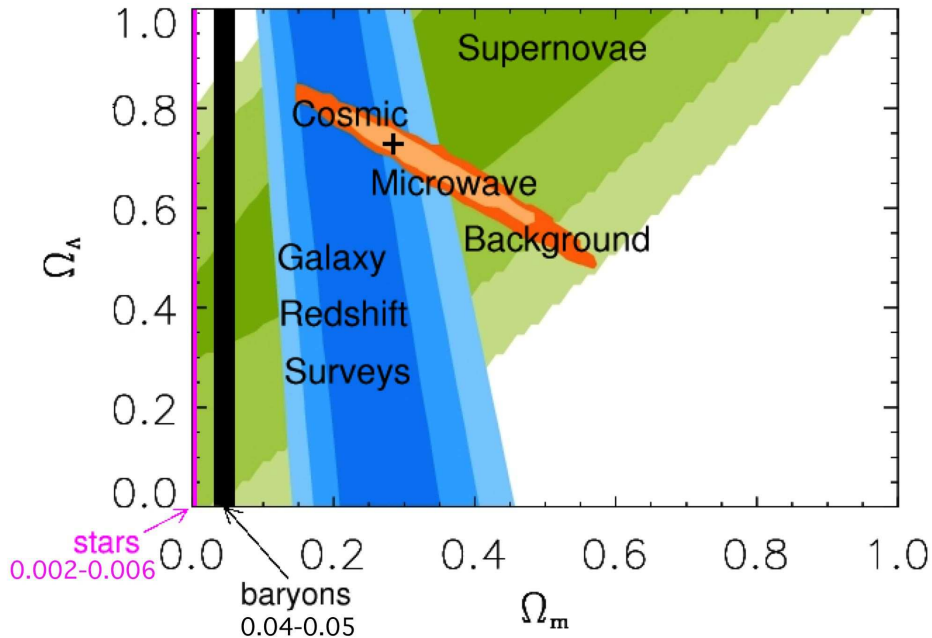}
\vspace{-12pt}
\caption{The concordance cosmology and the need for non-baryonic dark matter. Current cosmological measurements of the matter density $\Omega_{\rm m}$ and energy density $\Omega_{\Lambda}$ give the value marked with a cross at $\Omega_{\rm m} \simeq 0.27$, $\Omega_{\Lambda} \simeq 0.73$. The baryon density does not exceed 0.05 (black vertical band). The rest of the matter is non-baryonic. (Figure adapted from \citealp{Verde:2002}.)}
\end{figure}

A summary of the current measurements of the matter density $\Omega_{\rm m}$ and the energy density $\Omega_\Lambda$ are  shown in Figure~1 (adapted from \citealp{Verde:2002}). Both are in units of the critical density $\rho_{\rm crit} = 3H_0^2/(8\pi G)$, where $G$ is the Newton's gravitational constant, and $H_0$ is the present value of the Hubble constant. Three types of observations -- supernova measurements of the recent expansion history of the Universe, cosmic microwave background measurements of the degree of spatial flatness, and measurements of the amount of matter in galaxy structures obtained through big galaxy redshift surveys   --  agree with each other in a region around  the best current values of the matter and energy densities $\Omega_{\rm m} \simeq 0.27$ and $\Omega_{\Lambda} \simeq 0.73$ (cross in Figure~1). Measurements of the baryon density in the Universe using the cosmic microwave background spectrum and primordial nucleosynthesis constrain the baryon density $\Omega_{\rm b}$ to a value less than $\sim 0.05$ (black vertical band in Figure~1). The difference $ \Omega_{\rm m}  - \Omega_{\rm b} \simeq 0.22$ must be in the form of non-baryonic dark matter.\footnote{The red vertical band labeled `stars' in Figure~1 shows the density of luminous matter, corrected for all expected dim stars and gas \citep[see, e.g.,][]{Fukugita:1998}. The difference between the amount of luminous matter and the amount of baryons constitutes the dark baryon problem, which will not be addressed here \citep[see, e.g.,][]{Silk:2003}.} 

A precise determination of the cosmological density parameters is able to give the matter and energy densities in physical units. For example, in units of $1.879\times 10^{-29}$ g/cm$^3$ = 18.79 yg/m$^3$, \citet{Spergel:2003} have determined a total matter density 
\begin{equation}
\Omega_{\rm m} h^2 = 0.135 ^{+0.008}_{-0.009} ,
\end{equation}
of which 
\begin{equation}
\label{eq:omeganu}
\Omega_{\nu} h^2 < 0.0076
\end{equation}
is in the form of neutrinos (to 95\% confidence level),
\begin{equation}
\Omega_{\rm b} h^2 = 0.0224 \pm 0.0009
\end{equation}
is in the form of baryons (protons and nucleons in cosmological parlance),
and
\begin{equation}
\label{eq:omegacdm}
\Omega_{\rm CDM} h^2 = 0.113 ^{+0.008}_{-0.009}
\end{equation}
is in the form of cold dark matter (CDM), a non-baryonic component whose nature we are still trying to uncover. Some ideas of what it may be are presented in the next Section. 

\section{Popular candidates for non-baryonic dark matter}

A new kind of elementary particle has been the dominant (exclusive?) candidate for non-baryonic dark matter. 

A major classification of non-baryonic dark matter is based on its temperature at the time of galaxy formation, which occurs at a photon temperature of about 1 keV. Hot dark matter was relativistic at the time of galaxy formation, and as a consequence hindered the formation of the smallest objects by streaming out of the forming structures. An example of a hot dark matter particle is a light neutrino, much lighter than $\sim$keV. Cold dark matter was non-relativistic when galaxies formed, and thus was able to collapse effectively under the action of gravity because of its negligible pressure. Examples of cold dark matter particles are neutralinos, axions, WIMPZILLAs,  solitons (B-balls and Q-balls), etc. Warm dark matter was semi-relativistic at the time of galaxy formation, and is therefore an intermediate case between hot and cold dark matter. Two examples of warm dark matter are keV-mass sterile neutrinos and gravitinos.

Another important classification of particle dark matter rests upon its production mechanism. Particles that were in thermal equilibrium in the early Universe, like neutrinos, neutralinos, and most other WIMPs (weakly interacting massive particles), are called thermal relics. Particles which were produced by a non-thermal mechanism and that never had the chance of reaching thermal equilibrium in the early Universe are called non-thermal relics. There are several examples of non-thermal relics: axions emitted by cosmic strings, solitons produced in phase transitions, WIMPZILLAs produced gravitationally at the end of inflation, etc.

For the sake of presentation, we find still another classification useful. We will divide candidates for particle dark matter into three categories: Type Ia, Type Ib, and Type II (following a common practice in superconductors and supernovas).
Type Ia candidates are those known to exist, foremost among them are the neutrinos.
Type Ib candidates are candidates which are still undiscovered but are `well-motivated.' By this we mean that (1) they have been proposed to solve genuine particle physics problems, a priori unrelated to dark matter, and
(2) they have interactions and masses specified within a well-defined (and consistent) particle physics model. We are aware of the arbitrariness of this classification, and reserve the honor of belonging to the Type Ib category only 
to a sterile neutrino, the axion, and the lightest supersymmetric particle (which may be a neutralino, a gravitino, or a sneutrino) Finally,
Type II candidates are all other candidates, some of which are examples of maybe fruitful ideas, such as WIMPZILLAs, solitons (B-balls, Q-balls), dark matter from extra-dimensions, self-interacting dark matter, string-inspired dark matter, string-perspired dark matter, etc. It goes without saying that a candidate may move up from Type II to Type Ib and even to Type Ia as our understanding of particle physics models progresses.

We now examine some of the current candidates. 

\subsection{Type Ia: candidates that exist}

Dark matter candidates that are known to exist in Nature have an obvious  advantage over candidates that have not been detected. The chief particles in this category are the neutrinos.

There are three known `flavors' of neutrinos: the electron neutrino $\nu_e$, the muon neutrino $\nu_\mu$, and the tau neutrino $\nu_\tau$. They are so named because they are produced or destroyed in concomitance with the electron, the muon, and the tau lepton, respectively. 

"If neutrinos had a mass, they would be a good candidate for the dark matter," said Steven Hawking. We now know that neutrinos, or at least some of the neutrinos, do have a mass. This was discovered indirectly through the observation of neutrino flavor oscillations, i.e.\ the spontaneous conversion of one neutrino flavor into another as a neutrino propagates from point to point. The connection between flavor oscillations and neutrino masses can be seen as follows.

Consider for simplicity two flavors of neutrinos instead of three, $\nu_e$ and $\nu_\mu$, say. Weak interactions produce the flavor eigenstates $|\nu_e\rangle$ and $|\nu_\mu\rangle$, which are associated with their respective charged leptons. However, these flavor eigenstates are {\it not} energy eigenstates. Let $|\nu_1\rangle$ and $|\nu_2\rangle$ denote the two energy eigenstates for the two-flavor system, with energies $E_1$ and $E_2$ respectively. Then the flavor and the energy eigenstates are connected by a unitary transformation,
\begin{equation}
\label{eq:5}
\left\{ \begin{array}{l} \displaystyle 
|\nu_e\rangle = \cos \theta \, |\nu_1\rangle + \sin \theta \, |\nu_2\rangle ,\\
\vspace{-7pt}\\
\displaystyle
|\nu_\mu\rangle = -\sin\theta \, |\nu_1\rangle + \cos \theta \, |\nu_2 \rangle .
\end{array} \right.
\end{equation}
Imagine that at time $t=0$ we produce a $\nu_e$, so that the initial wave function is 
\begin{equation}
| \psi(0) \rangle = | \nu_e \rangle .
\end{equation}
After a time $t$, the wave function evolves according to the system Hamiltonian $\hat{H}$ as (we use natural units $\hbar=c=1$)
\begin{equation}
| \psi(t) \rangle = e^{-i\hat{H}t} |\psi(0)\rangle = e^{-i\hat{H}t} |\nu_e\rangle .
\end{equation}
To see the evolution explicitly, we expand $|\nu_e\rangle$ into energy eigenstates, and obtain
\begin{equation}
| \psi(t) \rangle = \cos\theta e^{-iE_1 t} | \nu_1\rangle + \sin\theta e^{-iE_2 t} | \nu_2\rangle .
\end{equation}
We can now ask what is the probability of observing the neutrino in the state $|\nu_\mu\rangle$ after a time $t$, i.e.\ of observing a neutrino with flavor $\nu_\mu$ instead of the initial $\nu_e$. According to standard rules of quantum mechanics, this probability is
\begin{eqnarray}
{\rm Prob}(\nu_e\to\nu_\mu) & = & \big\vert \langle \nu_\mu | \psi(t) \rangle \big\vert^2 \\
& = & \big\vert \cos\theta e^{-iE_1 t} \langle \nu_\mu | \nu_1\rangle + \sin\theta e^{-iE_2 t} \langle \nu_\mu | \nu_2\rangle \big\vert^2 \\
& = & \big\vert - \cos\theta \sin\theta e^{-iE_1 t} + \sin\theta \cos\theta e^{-iE_2 t}  \big\vert^2 \\
& = & \sin^2\!2\theta \, \sin^2\!\left[\tfrac{1}{2}(E_2-E_1)t \right],
\end{eqnarray}
where we have used Eq.~(\ref{eq:5}) and $ \langle \nu_1 | \nu_1 \rangle = \langle \nu_2 | \nu_2 \rangle =1$, $  \langle \nu_1 | \nu_2 \rangle =0$.
For free relativistic neutrinos, with momentum $p$ much larger than their mass $m$, we have
\begin{equation}
E = \sqrt{p^2+m^2} \simeq p + \frac{m^2}{2p} ,
\end{equation}
and
\begin{equation}
E_2-E_1 \simeq \frac{ m_2^2-m_1^2}{2p} .
\end{equation}
Hence
\begin{equation}
{\rm Prob}(\nu_e\to\nu_\mu) = \sin^2\!2\theta \, \sin^2\!\left[ \frac{ (m_2^2-m_1^2) t}{4 p} \right] .
\end{equation}
This equation shows that the probability of conversion from flavor $\nu_e$ to flavor $\nu_\mu$ oscillates in time with a frequency proportional to the difference of the squares of the neutrino masses $\Delta m_{12}^2 = m_2^2-m_1^2$. The observation of neutrino flavor oscillations therefore implies that neutrino masses differ from each other, and in particular that at least one of them is different from zero.

Neutrino oscillations have up to now been detected in two systems. Atmospheric muon neutrinos, which originate from the collision of cosmic rays with the Earth atmosphere, have been observed to oscillate into tau neutrinos \citep{Fukuda:1998}, 
\begin{equation}
\label{eq:deltamsquare1}
\nu_\mu \to \nu_\tau, \qquad \Delta m_{23}^2 \sim 3 \times 10^{-3} {\rm ~eV}^2.
\end{equation}
Solar neutrinos, produced in the nuclear reactions that make the Sun shine, also show oscillations \citep{SNO:2003},
\begin{equation}
\label{eq:deltamsquare2}
\nu_e \to \nu_\mu {\rm ~or~} \nu_\tau, \qquad \Delta m_{12}^2 \sim 7 \times 10^{-5} {\rm ~eV}^2 .
\end{equation}

These results can be used to set a lower limit on the mass of the heaviest neutrino.  Indeed, the mass of the heaviest neutrino must be greater than or equal to the square root of the largest mass-squared difference (just take the mass of the other neutrino to vanish). This gives the lower limit
\begin{equation}
\label{eq:loweroscill}
{\rm mass~of~heaviest~neutrino} \gtrsim 0.05 {\rm ~eV}.
\end{equation}

Upper limits on neutrino masses come from laboratory experiments, such as tritium decay and high-energy accelerator experiments, and are \citep[see][]{RPP}
\begin{equation}
\label{eq:upperacc}
m_1 < 2.8 {\rm ~eV}, \qquad m_2 < 190 {\rm ~keV}, \qquad m_3 < 18.2 {\rm ~MeV} .
\end{equation}
However, the small mass differences implied by Eqs.~(\ref{eq:deltamsquare1}) and~(\ref{eq:deltamsquare2}) imply that the smallest of the three upper limits applies to all three active neutrino masses. Thus we have
\begin{equation}
m_i < 2.8 {\rm ~eV} \qquad (i=1,2,3).
\end{equation}

It follows from this mass constraint that reactions such as $\nu_e \overline{\nu}_e \leftrightarrow e^+e^-$ in the hot early universe were able to keep standard-model neutrinos in thermal equilibrium. The neutrino density then follows from a computation of the neutrino number density (see Section 2.2). The result is
\begin{equation}
\label{eq:omegalightnu}
\Omega_\nu h^2 = \sum_{i=1}^3  \frac{g_i m_i}{90 {\rm ~eV}} ,
\end{equation}
where $g_i=1$ for a neutrino which is its own antiparticle (Majorana neutrino) and $g_i=2$ for a neutrino which is not its own antiparticle (Dirac neutrino). 

We already mentioned in Section 1 that cosmology provides an upper limit on the neutrino density $\Omega_\nu h^2$. This translates into a cosmological upper limit on the neutrino mass using Eq.~(\ref{eq:omegalightnu}). The cosmological limit is strictly speaking on the mass density in relativistic particles at the time of galaxy formation. An excessive amount of relativistic particles when galaxies form, i.e.\ of particles with mass $m \ll {\rm keV}$, would erase too much structure at the smallest scales. A combination of cosmic microwave background measurements, galaxy clustering measurements, and observations of the Lyman-$\alpha$ forest gives the upper limit quoted before \citep{Spergel:2003} \begin{equation}
\label{eq:omeganulimit}
\Omega_{\nu} h^2 < 0.0076 \qquad (95\% {\rm ~C.L.}).
\end{equation}
Eq.~(\ref{eq:omegalightnu}) then gives
\begin{equation}
\label{eq:upperlightnu}
g_1 m_1 + g_2 m_2 + g_3 m_3 < 0.7 {\rm ~eV}.
\end{equation}

On the other hand, Eq.~(\ref{eq:omegalightnu}) can be used in conjunction with inequality (\ref{eq:loweroscill}) to obtain a lower bound on the cosmological density in neutrinos. Taking only one massive Majorana flavor,
\begin{equation}
\Omega_\nu h^2 \gtrsim 0.0006.
\end{equation}
Thus neutrinos are definitely a form of dark matter, although perhaps a minor component of it.

The results for the known neutrinos as dark matter can be summarized by the constraints
\begin{eqnarray}
& 0.05 {\rm ~eV} < m_1+m_2+m_3 < 0.7 {\rm ~eV} , & \\
& 0.0006 < \Omega_\nu h^2 < 0.0076 , &
\end{eqnarray}
where the constraint in Eq.~(\ref{eq:upperlightnu}) has somewhat been relaxed by taking $g_i=1$. 

The upper limit on $\Omega_{\nu} h^2$ forbids currently known neutrinos from being the major constituents of dark matter. Moreover, since they are light and relativistic at the time of galaxy formation, the three neutrinos known to exist are hot, not cold, dark matter. 

The three active neutrinos are our only known particle candidates for non-baryonic dark matter. Since they fail to be cold dark matter, we are lead to consider hypothetical particles.

\subsection{Type Ib: `well-motivated' candidates}

We will discuss three cold dark matter candidates which are `well-motivated', i.e.\ that have been proposed to solve problems in principle unrelated to dark matter and whose properties can be computed within a well-defined particle physics model. The three candidates we discuss are: (1) a heavy active neutrino with standard model interactions, (2) the neutralino in the minimal supersymmetric standard model, and (3) the axion. Examples of other candidates that can be included in this category are a sterile neutrino \citep[see, e.g.,][]{Fuller} and other supersymmetric particles such as the gravitino \citep[see, e.g.,][]{Ellis:1983} and the sneutrino \citep[see, e.g.,][]{Murayama}.

The first two candidates we discuss belong to a general class called weakly interacting massive particles (WIMPs).\footnote{Notice that according to the Merriam-Webster Dictionary of the English Language, a wimp is a weak, cowardly, or ineffectual person.} WIMPs that were in thermal equilibrium in the early universe (thermal WIMPs) are particularly interesting. Their cosmological density is naturally of the right order of magnitude when their interaction cross section is of the order of a weak cross section. This also makes them detectable in the laboratory, as we will see later. In the early Universe, annihilation reactions that convert WIMPs into standard model particles were initially in equilibrium with their opposite reactions. As the universe expanded, and the temperature became smaller than the WIMP mass, the gas of WIMPs, still in equilibrium, diluted faster than the gas of standard model particles. This occurred because the equilibrium number density of non-relativistic particles is suppressed by a Boltzmann factor $e^{-m/T}$ with respect to the number density of relativistic particles. After a while, WIMPs became so rare that the WIMP annihilation reactions could no longer occur (chemical decoupling), and from then on the number density of WIMPs decreased inversely with volume (or in other words, the number of WIMPs per comoving volume remained constant). Chemical decoupling occurs approximately when the WIMP annihilation rate $\Gamma_{\rm ann} = \langle \sigma_{\rm ann} v \rangle n $ became smaller than the universe expansion rate $H$. Here $\sigma_{\rm ann}$ is the WIMP annihilation cross section, $v$ is the relative velocity of the annihilating WIMPs, $n$ is the WIMP number density, and the angle brackets denote an average over the WIMP thermal distribution. Using Friedmann's equation to find the expansion rate $H$ gives
\begin{equation}
\label{eq:omegawimp}
\Omega h^2 \approx \frac{ 3 \times 10^{-27} {\rm ~cm^3/s} }{ \langle \sigma_{\rm ann} v \rangle } 
\end{equation}
for the relic density of a thermal WIMP. An important property of this equation is that smaller annihilation cross sections correspond to larger relic densities (``The weakest wins.'') This can be understood from the fact that WIMPs with stronger interactions remain in chemical equilibrium for a longer time, and hence decouple when the universe is colder, wherefore their density is further suppressed by a smaller Boltzmann factor. Figure 2 illustrates this relationship.

\begin{figure}[!b]
\label{fig:2}
\begin{minipage}{0.6\textwidth}
\includegraphics[width=0.9\textwidth]{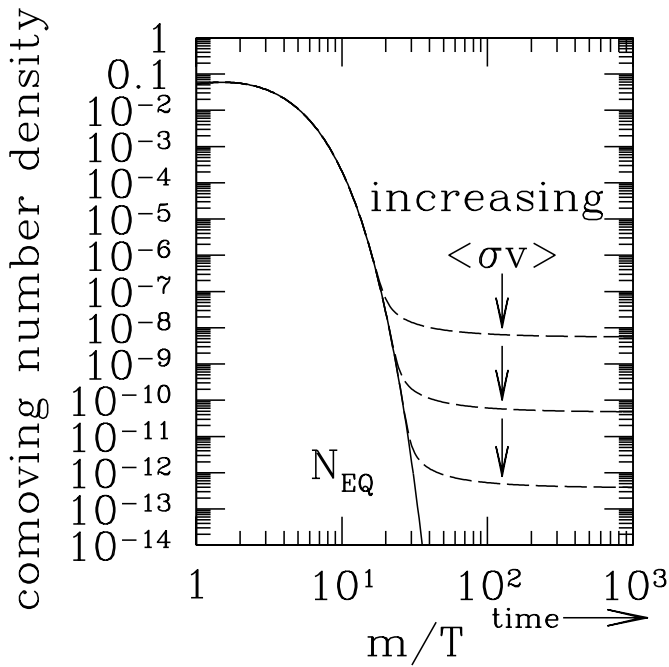}
\end{minipage}
\begin{minipage}{0.4\textwidth}
\caption{Evolution of a typical WIMP number density in the early universe. The number of WIMPs in a volume expanding with the universe (comoving density) first decreases exponentially due the Boltzmann factor $e^{-m/T}$ and then `freezes out' to a constant value when the WIMP annihilation reactions cannot maintain chemical equilibrium between WIMPs and standard model particles. In the figure, $\langle \sigma v \rangle $ is the thermally averaged annihilation cross section times relative velocity. WIMPs with larger annihilation cross section end up with smaller densities.} 
\end{minipage}
\end{figure}

It must be remarked here that in the non-relativistic limit $v\to0$, the product $\sigma_{\rm ann} v$ tends to a constant, because the annihilation cross section $\sigma_{\rm ann}$ diverges as $1/v$ as $v\to0$. This is analogous to what happens for the scattering cross section of thermal neutrons.

\subsubsection*{Heavy neutrino}

The WIMP {\it par excellence} is a heavy neutrino. The example we consider is a thermal Dirac neutrino $\nu$ of the fourth generation with Standard Model interactions and no lepton asymmetry. Figure 3 summarizes its relic density as a function of mass. Also shown in the Figure are the current constraints from accelerator experiments and dark matter searches. 

A neutrino lighter than $\sim1$ MeV decouples while relativistic. If it is so light to be still relativistic today ($m_\nu \lesssim 0.1 {\rm ~meV}$), its relic density is $\rho_\nu = 7\pi^2 T_\nu^4/120$. If it became non-relativistic after decoupling, its relic density is determined by its equilibrium number density as $\rho_\nu=m_\nu 3\zeta(3)T_\nu^3/2\pi^2$. Here $T_\nu = (3/11)^{1/3} T_\gamma$, where $T_\gamma = 2.725 \pm 0.002 {\rm K}$ is the cosmic microwave background temperature. (We use natural units, $c=\hbar=1$.)

A neutrino heavier than $\sim1$ MeV decouples while non-relativistic. Its relic density is determined by its annihilation cross section, as for a general WIMP (see Eq.~(\ref{eq:omegawimp})). The shape of the relic density curve in Figure 3 is a reflection of the behavior of the annihilation cross section. The latter is dominated by the Z-boson resonance at $m_\nu \simeq  m_Z/2$. This resonant annihilation gives the characteristic V shape to the relic density curve. Above $m_\nu \sim 100 {\rm ~GeV}$, new annihilation channels open up, namely the annihilation of two neutrinos into two Z- or W-bosons. The new channels increase the annihilation cross section and thus lower the neutrino relic density. Soon, however, the perturbative expansion of the cross section in powers of the (Yukawa) coupling constant becomes untrustworthy (the question mark in Figure 3). An alternative unitarity argument limits the Dirac neutrino relic density to the dashed curve on the right in the Figure. Neutrinos heavier than 10 TeV `overclose' the universe, i.e.\ have a relic density that corresponds to a universe which is too young. 

The `dark matter' band in Figure 3 indicates where the neutrino is a good dark matter candidate (the band is actually quite generous in light of the most recent measurements of $\Omega h^2$). A thermal Dirac neutrino is a good dark matter candidate when its mass is around few eV, a few GeV or possibly a TeV.  For masses smaller than about an eV and between $\sim$10 GeV and $\sim$100 GeV, it is an underabundant relic from the Big Bang, too dilute to be a major component of the dark matter but nevertheless a cosmological relic. For other masses, it is cosmologically excluded.

\begin{figure}[t]
\label{fig:3}
\begin{minipage}{0.3\textwidth}
\caption{Neutrinos as dark matter. Relic density of a thermal Dirac neutrino with standard-model interactions, together with current constraints from cosmology, accelerators (LEP), and dark matter searches. See text for explanations. (The `dark matter' band is quite generous in light of the WMAP measurements.) }
\end{minipage}
\begin{minipage}{0.7\textwidth}
\includegraphics[width=\textwidth]{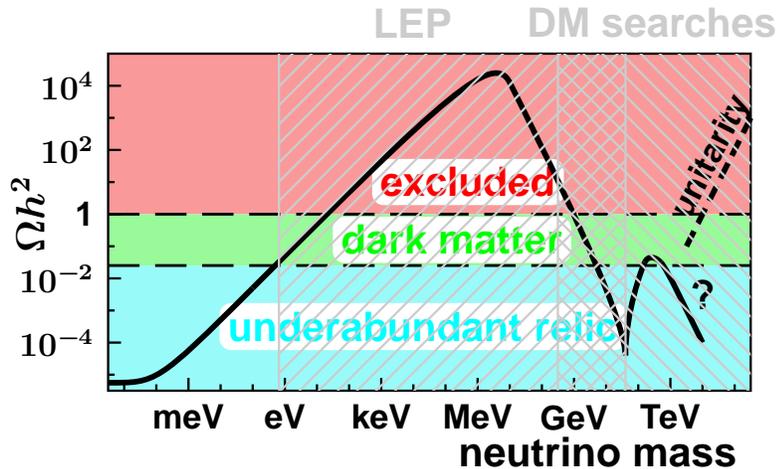}
\end{minipage}
\end{figure}

Dark matter neutrinos with a mass around 1 eV would be relativistic at the time of galaxy formation ($\sim$ keV), and would thus be part of hot dark matter. From the bounds on hot dark matter in the preceding Section, however, they cannot be a major component of the dark matter in the Universe. 

Neutrinos can be cold dark matter if their masses are around few GeV or a TeV. However,  fourth-generation heavy neutrinos lighter than 45 GeV are excluded by the measurement of the Z-boson decay width at the Large Electron-Positron collider at CERN. Moreover, direct searches for WIMP dark matter in our galaxy exclude Dirac neutrinos heavier than $\sim$0.5 GeV as the dominant component of the galactic dark halo (see Figure 3). Thus although heavy Dirac neutrinos could still be a tiny part of the halo dark matter, they cannot solve the cold dark matter problem. 

We need another non-baryonic candidate for cold dark matter.

\subsubsection*{Neutralino}

The WIMP {\it par default} is the lightest neutralino $\tilde{\chi}_1^0$, or sometimes simply $\chi$, which is often the lightest supersymmetric particle in supersymmetric extensions of the Standard Model of particle physics. Supersymmetry is a new symmetry of space-time that has been discovered in the process of unifying the fundamental forces of nature (electroweak, strong, and gravitational). Supersymmetry also helps in stabilizing the masses of fundamental scalar particles in the theory, such as the Higgs boson, a problem, called the hierarchy problem, which basically consists in explaining why gravity is so much weaker than the other forces. 

Of importance for cosmology is the fact that supersymmetry requires the existence of a new particle for each particle in the Standard Model. These supersymmetric partners differ by half a unit of spin, and come under the names of sleptons (partners of the leptons), squarks (partners of the quarks), gauginos (partners of the gauge bosons) and higgsinos (partners of the Higgs bosons). Sleptons and squarks have spin 0, and gauginos and higgsinos have spin $\tfrac{1}{2}$.

If supersymmetry would be an explicit symmetry of nature, superpartners would have the same mass as their corresponding Standard Model particle. However, no Standard Model particle has a superpartner of the same mass. It is therefore assumed that supersymmetry, much as the weak symmetry, is broken. Superpartners can then be  much heavier than their normal counterparts,  explaining why they have not been detected so far. However, the mechanism of supersymmetry breaking is not completely understood, and in practice it is implemented in the model by a set of supersymmetry-breaking parameters that govern the values of the superpartners masses (the superpartners couplings are fixed by supersymmetry). 

The scenario with the minimum number of particles is called the minimal supersymmetric standard model or MSSM. The MSSM has 106 parameters  beyond those in the Standard Model: 102 supersymmetry-breaking parameters, 1 complex supersymmetric parameter $\mu$, and 1 complex electroweak symmetry-breaking parameter $\tan\beta$ (see, e.g., the article by Haber in \citealp{RPP}). Since it is cumbersome to work with so many parameters, in practice phenomenological studies consider simplified scenarios with a drastically reduced number of parameters. The most studied case (not necessarily the one Nature has chosen) is minimal supergravity, which reduces the number of parameters to five: three real mass parameters at the Grand Unification scale (the scalar mass $m_0$, the scalar trilinear coupling $A_0$, and the gaugino mass $m_{1/2}$) and two real parameters at the weak scale (the ratio of Higgs expectation values $\tan \beta$ and the sign of the $\mu$ parameter). Other scenarios are possible and are considered in the literature. Of relevance to dark matter studies is, for example, a class of models with seven parameters specified at the weak scale: $\mu$, $\tan\beta$, the gaugino mass parameter $M_2$, the mass $m_A$ of the CP-odd Higgs boson, the sfermion mass parameter $\widetilde{m}$, the bottom and top quark trilinear couplings $A_b$ and $A_t$. See the reviews by \citet{Jungman:1996} and \citet{Bergstrom:2000} for more details.

It was realized long ago by \citet{Goldberg:1983} and \citet{Ellis:1983} that the lightest superposition of the neutral gauginos and the neutral higgsinos (which having the same quantum numbers mix together) is an excellent dark matter candidate. It is often the lightest supersymmetric particle, it is stable under the requirement that superpartners are only produced or destroyed in pairs (called R-parity conservation), it is weakly interacting, as dictated by supersymmetry, and it is massive. It is therefore a genuine WIMP, and it is among the most studied of the dark matter candidates. Its name is the lightest neutralino.

\begin{figure}[t]
\label{fig:4}
\begin{minipage}{0.65\textwidth}
\includegraphics[width=0.9\textwidth]{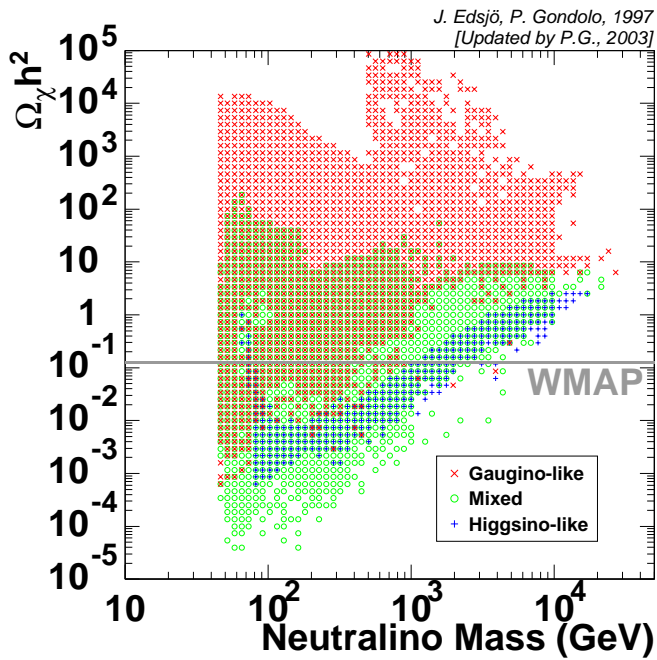}
\end{minipage}
\begin{minipage}{0.34\textwidth}
\caption{Relic density of the lightest neutralino as a function of its mass. For each mass, several density values are possible depending on the other supersymmetric parameters (seven in total in the scenario plotted). The color code shows the neutralino composition (gaugino, higgsino or mixed). The gray horizontal line is the current error band in the WMAP measurement of the cosmological cold dark matter density. (Figure adapted from \citealp{Edsjo:1997}.)}
\end{minipage}
\end{figure}

Several calculations exist of the density of the lightest neutralino. An example is given in Figure 4, which reproduces a figure from \citet{Edsjo:1997}, updated with the WMAP value of the cold dark matter density. This figure was obtained in a scenario with seven supersymmetric parameters at the weak scale. The relic density is not fixed once the neutralino mass is given, as was the case for a Dirac neutrino, because the neutralino annihilation cross section depends on the masses and composition of many other supersymmetric particles, thus ultimately on all supersymmetric parameters. Therefore the density in Figure 4 is not a single-valued function of the neutralino mass, and the plot must be obtained through an extended computer scan in the seven-dimensional parameter space.\footnote{We may ask if there can be points in the empty regions, or in more general terms, what is the meaning of the density of points in Figure 4, and in similar figures in Section 3. For a discussion on this, see \citet{Bergstrom:1996}.}
It is clear from Figure 4 that it is possible to choose the values of the supersymmetric parameters in a way that the neutralino relic density satisfies the current determination of $\Omega h^2$. Although it may seem ridiculous to claim that the neutralino is {\it naturally} a good dark matter candidate, let us notice that the neutralino relic density in Figure 4 and the neutrino relic density in Figure 3 have a similar range of variation. It is the precision of the cosmological measurements that make us think otherwise.

Reversing the argument, the precision of the cosmological measurements can be used to select the regions of supersymmetric parameter space where the lightest neutralino is cold dark matter. With the current precision of cosmological measurements, these are very thin regions in supersymmetric parameter space. For this approach to be carried out properly, the theoretical calculation of the neutralino relic density should match the precision of the cosmological data. The latter is currently 7\%, as can be gathered from Eq.~(\ref{eq:omegacdm}), and is expected to improve to about 1\% before the end of the decade with the launch of the Planck mission. A calculation of the neutralino relic density good to 1\% now exists, and is available in a computer package called DarkSUSY \citep{Gondolo:2000c,Gondolo:2002}. The 1\% precision refers to the calculation of the relic density starting from the supersymmetric parameters at the weak scale. The connection with the parameters at the Grand Unification scale, which is vital for minimal supergravity, introduces instead large errors in important regions of parameter space, errors that some authors estimate to be as big as 50\% \citep{Allanach:2003}. 

\begin{figure}[!t]
\label{fig:5}
\begin{minipage}{0.27\textwidth}
\caption{Illustration of the power of WMAP constraints on the minimal supergravity parameter space. The figure shows a slice in $m_0$ and $m_{1/2}$ with $\mu<0$ at $\tan\beta=30$ and $A_0=0$. The WMAP constraint is a very thin (grey) line that approximately follows the edges of the allowed region. (Figure adapted from \citealp{Edsjo:2003}.)}
\end{minipage}
\begin{minipage}{0.72\textwidth}
\includegraphics[width=1.05\textwidth]{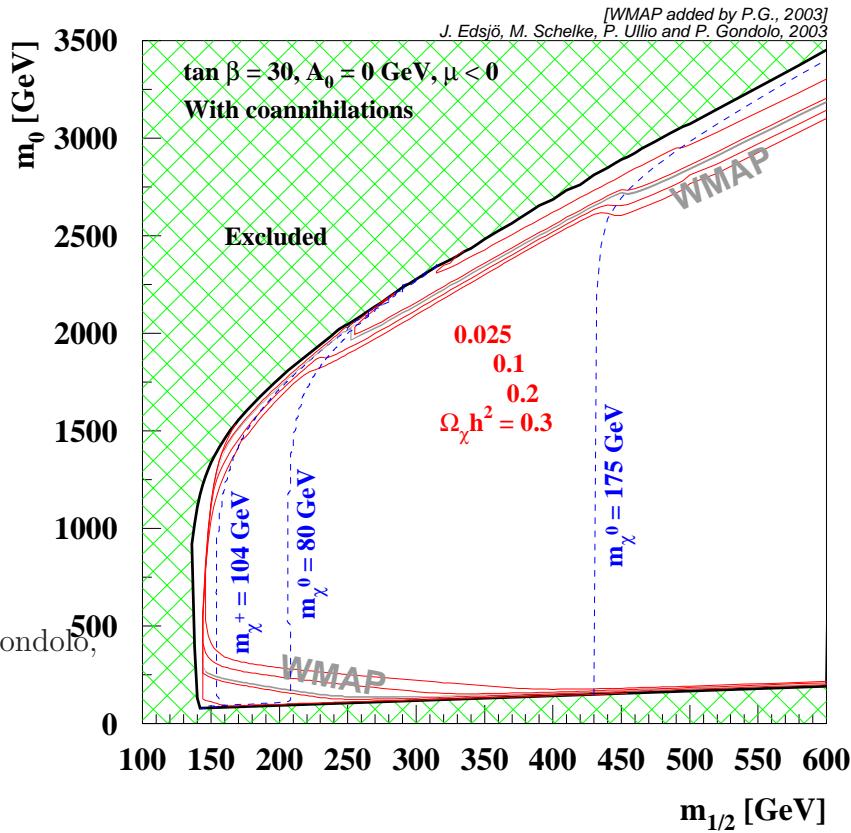}
\end{minipage}
\end{figure}

Given the importance of this calculation, we give now a rapid survey of the ingredients needed to achieve a precision of 1\%. Firstly, the equation governing the evolution of the neutralino number density $n$,
\begin{equation}
\dot n +  3 H n = - \langle \sigma_{\rm ann} v \rangle ( n^2 - n_{\rm eq}^2 ),
\end{equation}
 has to be solved numerically. The equation being `stiff' (i.e.\ its difference equation  being stable only for unreasonably small stepsizes), a special numerical method should be used. The thermal average of the annihilation cross section $  \langle \sigma_{\rm ann} v \rangle $ at temperature $T$ should be computed relativistically, since the typical speed of neutralinos at decoupling is of the order of the speed of light, $v \sim c/3$. For this purpose, we can use the expression in \citet{Gondolo:1991}:
 \begin{equation}
 \langle \sigma_{\rm ann} v \rangle = \frac{ \int_0^\infty dp \, p ^2 \, W(p) \, K_1(\sqrt{s}/T) } { m^4 T \left[ K_2(m/T) \right]^2 } ,
 \end{equation}
 where $W(p)$ is the annihilation rate per unit volume and unit time (a relativistic invariant), $s = 4 ( m^2 + p^2) $ is the center-of-mass energy squared, and $K_1$, $K_2$ are modified Bessel functions. In passing, let us remark that the common method of expanding in powers of $v^2$, $\sigma_{\rm ann} v = a + b v^2 + \cdots$, and then taking the thermal average to give $\langle \sigma_{\rm ann} v \rangle = a + b \tfrac{3T}{2m} + \cdots $ is unreliable, since it gives rise to negative $\langle \sigma_{\rm ann} v \rangle$, and thus negative $\Omega h^2$, near resonances and thresholds. These are nowadays the most important regions of parameter space. Finally, an essential ingredient in the calculation of the neutralino relic density is the inclusion of coannihilation processes. These are processes that deplete the number of neutralinos through a chain of reactions, and occur when another supersymmetric particle is close in mass to the lightest neutralino ($\Delta m \sim T$). In this case, scattering of the neutralino off a particle in the thermal `soup' can convert the neutralino into the other supersymmetric particle close in mass, given that the energy barrier that would otherwise have prevented it (i.e.\ the mass difference) is easily overcome.  The supersymmetric particle participating in the coannihilation may then decay and/or react with other particles and eventually effect the disappearance of neutralinos. We give two examples. Coannihilation with charginos $\tilde{\chi}^\pm$ (partners of the charged gauge and Higgs bosons) may proceed via, for instance,
\begin{equation}
\tilde{\chi}_1^0 e^- \to \tilde{\chi}_2^- \nu_e, \qquad \tilde{\chi}_2^- \to \tilde{\chi}_2^0 d \bar{u} , \qquad \tilde{\chi}_2^0 \tilde{\chi}_1^0 \to W^+ W^- 
\end{equation}
(subscripts on superpartner names indicate particles with different masses). Coannihilation with tau sleptons $\tilde{\tau}$ may instead involve the processes 
\begin{equation}
\tilde{\chi}_1^0 \tau \to \tilde{\tau} \gamma, \qquad \tilde{\tau} \tilde{\chi}_1^+ \to \tau W^+ .
\end{equation}
Coannihilations were first included in the study of near-degenerate heavy neutrinos by \citet{Binetruy:1984} and were brought to general attention by \citet{Griest:1991}. The current state-of-the-art treatment of neutralino coannihilations, which involves several thousands of processes, is contained in the work by \citet{Edsjo:1997} and \citet{Edsjo:2003}. 

To illustrate the power of the cosmological precision measurements in selecting regions of supersymmetric parameter space, Figure 5 shows  the WMAP constraint in one of the figures in \citet{Edsjo:2003}. The constraint is as a very thin line that approximately follows the edges of the allowed region in this slice of parameter space (the $m_0$--$m_{1/2}$ plane with $\mu<0$, $\tan\beta=30$ and $A_0=0$). 

Cosmological constraints on supersymmetric models are very powerful, and may even serve as a guidance in searching for supersymmetry. This partially justifies the extensive literature on the subject. The neutralino as dark matter is certainly `fashionable.'

\subsubsection*{Axion}

Our third and last example of a `well-motivated' cold dark matter candidate is the axion. 

Axions were suggested by \citet{Peccei} to solve the so-called ``strong CP problem''. Out of the vacuum structure of Quantum Chromodynamics there arises a large CP-violating phase, which is at variance with stringent measurements of the electric dipole moment of the neutron, for example. A possible solution to this problem is that the CP-violating phase is the vacuum expectation value of a new field, the axion, which relaxes dynamically to a very small value. The original axion model of Peccei and Quinn is today experimentally ruled out, but other axion models based on the same idea have been proposed. Among them are the invisible axions of \citet{Kim} and \citet{Shifman} (KSVZ axion) and of \citet{Dine} and \citet{Zhitnitsky} (DFSZ). They differ in the strength of the axion couplings to matter and radiation.

\begin{figure}[!t]
\label{fig:6}
\begin{minipage}{0.39\textwidth}
\caption{Laboratory, astrophysical, and cosmological constraints on the axion mass $m_A$. The inflation scenario and the string scenario are referred in the text as the vacuum alignment scenario and the string emission scenario, respectively. $f_A$ is the axion decay constant, which is inversely related to $m_A$. The axion is a good dark matter candidate for 1 $\mu$eV $\lesssim m_A \lesssim$ 1 meV. (Figure from Raffelt in \citealp{RPP}.)}
\end{minipage}
\begin{minipage}{0.6\textwidth}
\vspace{12pt}
\includegraphics[width=\textwidth]{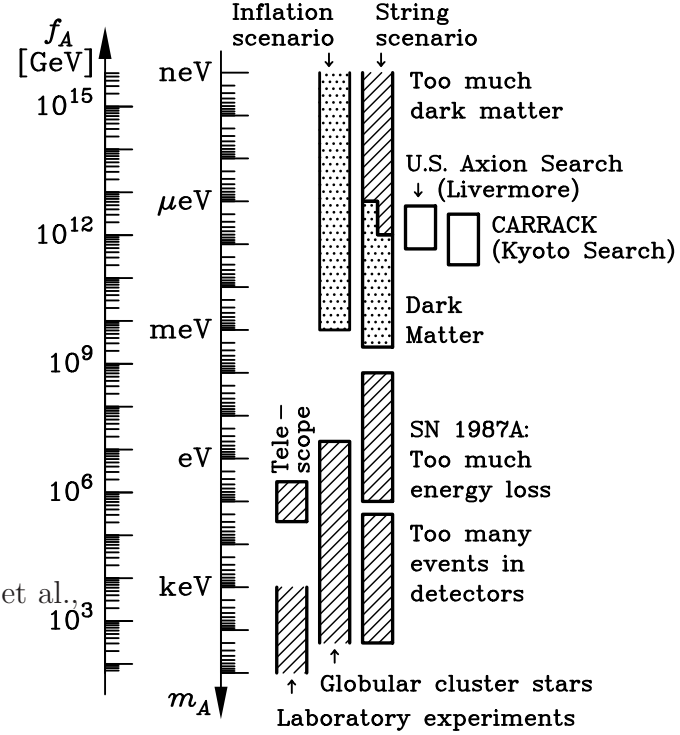}
\end{minipage}
\end{figure}

\begin{figure}[!b]
\label{fig:7}
\begin{minipage}{0.6\textwidth}
\includegraphics[width=1.1\textwidth]{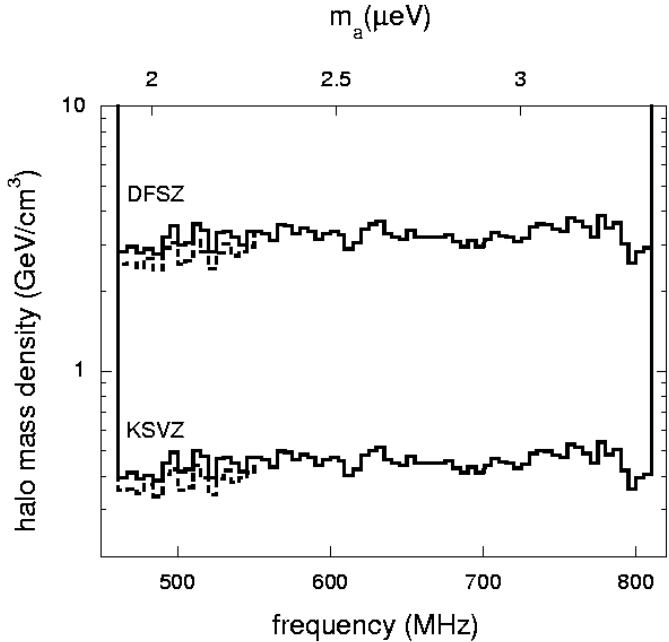}
\end{minipage}
\begin{minipage}{0.39\textwidth}
\caption{Experimental constraints on the density of axions in the galactic halo near the Sun as a function of the axion mass (upper scale) and cavity frequency (lower scale). The regions above the curves marked `DFSZ' and `KSVZ' are excluded fro the respective axion models. The currently accepted value for the local dark halo density is 0.45 GeV/cm$^3$, which is approximately the extension of the excluded region for the KSVZ axion. (Figure from \citealp{Asztalos:2004}.) }
\end{minipage}
\end{figure}

In a cosmological context, axions, contrary to neutrinos and neutralinos, are generally produced non-thermally (although thermal axion production is sometimes considered, as are non-thermal neutrino and neutralino productions). The two main mechanisms for non-thermal axion production are vacuum alignment and emission from cosmic strings. In the vacuum alignment mechanism, a potential is generated for the axion field at the chiral symmetry breaking, and the axion field, which can in principle be at any point in this potential, starts moving toward the minimum of the potential and then oscillates around it. Quantum-mechanically, the field oscillations correspond to the generation of axion particles. In the other main non-thermal mechanism for axion production, axions are emitted in the wiggling or decay of cosmic strings. In both cases, axions are produced with small momentum, $\ll$ keV, and thus they are cold dark matter despite having tiny masses, between 1 $\mu$eV and 1 meV. This is in fact the range of masses in which axions are good dark matter candidates. Figure 6 shows the current constraints on the axion mass from laboratory, astrophysical, and cosmological data. 

Searches for axions as galactic dark matter rely on the coupling of axions to two photons. An incoming galactic axion can become a photon  in the magnetic field in a resonant cavity. For this to happen, the characteristic frequency of the cavity has to match the axion mass. Since the latter is unknown, searches for galactic axions use tunable cavities, and scan over the cavity frequency, a time-consuming process. The U.S. axion search at Livermore is currently exploring a wide range of interesting axion masses, and has put some constraint on the KSVZ axion as a dominant component of the galactic halo \citep{Asztalos:2004}. Figure 7 shows the constraints on the local galactic density in axions as a function of the axion mass. 
KSVZ axions with mass in the range 1.91-3.34 $\mu$eV cannot be the main component of galactic dark matter. The Livermore search is still continuing to a larger range of axion masses. It is fair to say that axion dark matter is either about to be detected or about to be ruled out.

\subsection{Type II: other candidates}

In the Type II category we put all hypothetical cold dark matter candidates that are neither Type Ia nor Type Ib. Some of these candidates have been proposed for no other reason than to solve the dark matter problem. Others are examples of beautiful ideas and clever mechanisms that can provide good possibilities for non-baryonic dark matter, but in some way or another lack the completeness of the theoretical particle physics models of Types Ia and Ib. Although Type II candidates are not studied as deeply as others, it may well be that eventually the question of the nature of cold dark matter might find its answer among them.

Below we present the idea of self-interacting dark matter and gravitationally-produced WIMPZILLAs. Other interesting candidates have been proposed recently in models with extra dimensions, such as Kaluza-Klein dark matter \citep{Cheng:2002} and branons \citep{branons:2003}. 

There are several ways in which one may be able to come up with an {\it ad hoc} candidate for non-baryonic cold dark matter. A humorous flowchart on how to do this was put together around 1986 by a group of graduate students at Princeton \citep{Lauer:1986}. The flowchart involves multiple options, and one possibility runs as follows. ``A new particle is envisioned which is cooked up just to make everything OK but violates federal law \ldots still, it doesn't prevent a paper being written by Spergel \ldots" We have now the proper setting to introduce a candidate suggested by \citet{Spergel:2000}, self-interacting dark matter.

\subsubsection*{Self-interacting dark matter}

The idea behind the introduction of self-interacting dark matter is to find a solution to the cusp and satellite problems of standard cold dark matter scenarios. These two problems are in effect discrepancies between observations and results of simulations of structure formation on the galactic scale. Namely, in the cusp problem, numerical simulations predict a dark matter density profile which increases toward the center of a galaxy like a power law $r^{-\gamma}$ with $\gamma \sim 1$ or higher. This sharp density increase is called a cusp. On the other hand, kinematical and dynamical determinations of the dark matter profile in the central regions of galaxies, especially of low surface brightness galaxies, tend not to show such a sharp increase but rather a constant density core. The observational situation is still rather confused, with some galaxies profiles being compatible with a cusp and others with a core. The theoretical situation is also not fully delineated, with higher resolution simulations showing a dependence of the slope $\gamma$ on the mass of the galaxy. Although many ideas have been proposed for the resolution of the cusp problem, it is still not completely resolved.

Perhaps connected with the cusp problem is the satellite problem, which is a mismatch between the observed and the simulated numbers of satellites in a galaxy halo. Too many satellites are predicted by the simulations. In reality, observations can detect only the luminous satellites while simulations contain all satellites, including the dark ones. It may be that many dark satellites do not shine, thus solving the satellite problem, but how and which satellites become luminous is not understood yet. 

\citet{Spergel:2000} suggested another way to solve both problems. They realized that if dark matter particles would interact with each other with a mean free path of the order of the size of galactic cores, the dark matter interactions would efficiently thermalize the system and avoid the formation of both a central cusp and too many satellites. The requirement on the mean free path $\lambda$ is roughly $\lambda \sim 10 $ kpc. We can figure out the necessary cross section $\sigma$ for dark matter self-interactions by recalling that the mean free path is related to the cross section and the number density $n$, or matter density $\rho=mn$, through the relationship $\lambda = 1/(n\sigma)  = m/(\rho\sigma)$. Taking a typical $\rho \sim 0.3$ GeV/cm$^3$ gives $\sigma/m \sim 60 $ cm$^2$/g. So \citet{Spergel:2000} suggested a new self-interacting dark matter particle with $\sigma/m$ in the range
\begin{equation}
1 {\rm ~cm^2/g} \lesssim \sigma/m \lesssim 100 {\rm ~cm^2/g} .
\end{equation}
To understand the magnitude of this number, it is useful to compare it with the geometric cross section of a proton, which is one of the largest known cross sections for elementary particles. We have
\begin{equation}
(\sigma/m)_{\rm proton} \sim 1 {\rm ~fm^2/GeV} \sim 0.006 {\rm ~cm^2/g}.
\end{equation}
Thus the desired $\sigma/m$ seems rather big. It is therefore not surprising that astrophysical constraints on self-interacting dark matter are rather stringent. \citet{Gnedin:2001} considered the evaporation of halos inside clusters and set the constraint $\sigma/m < 0.3 {\rm -} 1$ cm$^2$/g. \citet{Yoshida:2000} considered the shape of cluster cores, which is rounder for self-interacting dark matter than for standard cold dark matter, and concluded that $\sigma/m$ must be $ < 10$ cm$^2$/g. This bound was later strengthened by \citet{MiraldaEscude:2002} to $\sigma/m < 0.02$ cm$^2$/g. \citet{Markevitch:2003} discovered a gas bullet lagging behind dark matter in the merging galaxy cluster 1E0657-56. They combined Chandra X-ray maps of the hot gas in the cluster with weak lensing maps of its mass distribution. From estimates of the column mass densities and of the distance between the gas and the dark matter, Markevitch et al.\ were able to set the upper limit $\sigma/m < 10 $ cm$^2$/g with direct observations of the dark matter distribution. All these bounds leave little room, if any, to self-interacting dark matter a la Spergel \& Steinhardt.

\subsubsection*{WIMPZILLAs}

Our last example of cold dark matter candidates illustrates a fascinating idea for generating matter in the expanding universe: the gravitational creation of matter in an accelerated expansion. This mechanism is analogous to the production of Hawking radiation around a black hole, and of Unruh radiation in an accelerated reference frame.

WIMPZILLAs \citep{Chung:1998,Chung:1999,Kuzmin:1998} are very massive relics from the Big Bang, which can be the dark matter in the universe if their mass is $\approx 10^{13}$ GeV. They were produced at the end of inflation through a variety of possible mechanisms: gravitationally, during preheating, during reheating, in bubble collisions. It is possible that their relic abundance does not depend on their interaction strength but only on their mass, giving great freedom in their phenomenology. To be the dark matter today, they are assumed to be stable or to have a lifetime of the order of the age of the universe. In the latter case, their decay products may give rise to the highest energy cosmic rays, and solve the problem of cosmic rays beyond the GZK cutoff.

\begin{figure}[t]
\label{fig:8}
\centering
\includegraphics[width=0.8\textwidth]{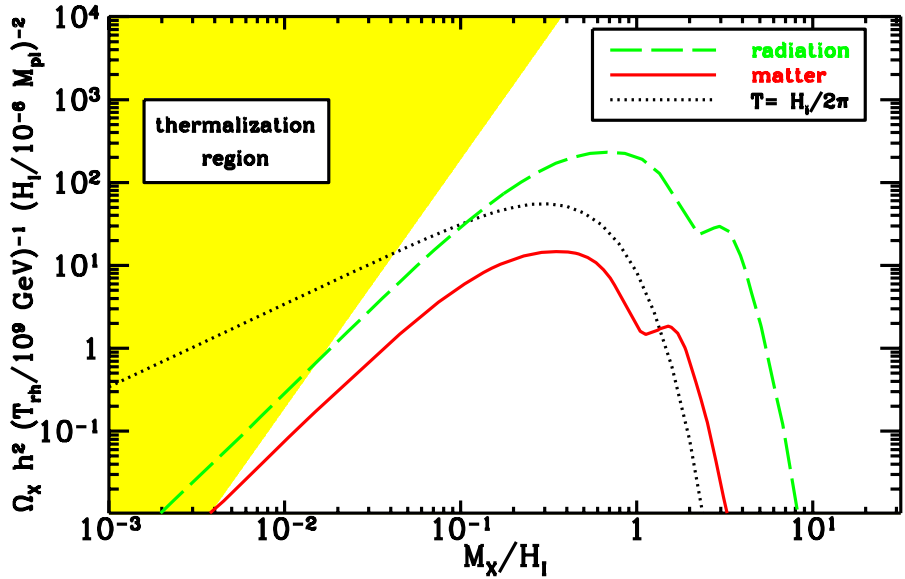}
\vspace{-12pt}
\caption{Relic density of gravitationally-produced WIMPZILLAs as a function of their mass $M_X$. $H_I$ is the Hubble parameter at the end of inflation, $T_{\rm rh}$ is the reheating temperature, and $M_{\rm pl} \approx 3 \times 10^{19} {\rm ~GeV}$ is the Planck mass. The dashed  and solid lines correspond to  inflationary models that smoothly end into a radiation  or matter dominated epoch, respectively. The dotted line is a thermal distribution at the temperature indicated. Outside the `thermalization region'  WIMPZILLAs cannot reach thermal equilibrium.
(Figure from \citealp{Chung:1998}.)}
\end{figure}

Gravitational production of particles is an important phenomenon that is worth describing here. Consider a scalar field (particle) $X$ of mass $M_X$ in the expanding universe. Let $\eta$ be the conformal time and $a(\eta)$ the time dependence of the expansion scale factor. Assume for simplicity that the universe is flat. The scalar field $X$ can be expanded in spatial Fourier modes as 
\begin{equation}
X(\vec{x},\eta) = \int \frac{ d^3 k}{ (2\pi)^{3/2} a(\eta) } \left[ a_k h_k(\eta) e^{i \vec{k} \cdot \vec{x} } + a_k^\dagger h_k^*(\eta) e^{-i \vec{k} \cdot \vec{x}} \right] .
\end{equation}
Here $a_k$ and $a_k^\dagger$ are creation and annihilation operators, and $h_k(\eta)$ are mode functions that satisfy (a) the normalization condition 
$
h_k h_k^{\prime *} - h'_k h_k^* = i
$
(a prime indicates a derivative with respect to conformal time),
and (b) the mode equation
\begin{equation}
\label{eq:modeequation}
h''_k(\eta) + \omega_k^2(\eta) \,  h_k(\eta) = 0 ,
\end{equation}
where
\begin{equation}
\omega_k^2(\eta) =  k^2 + M_X^2 a^2 + (6 \xi-1) \frac{a''}{a} .
\end{equation}
The parameter $\xi$ is $\xi=0$ for a minimally-coupled field and $\xi=\tfrac{1}{6}$ for a conformally-coupled field. The mode equation, Eq.~(\ref{eq:modeequation}), is formally the same as the equation of motion of a harmonic oscillator with time-varying frequency $\omega_k(\eta)$. For a given positive-frequency solution $h_k(\eta)$, the vacuum $ | 0_{h} \rangle$ of the field $X$, i.e.\ the state with no $X$ particles, is defined as the state that satisfies $a_k | 0_{h} \rangle = 0$ for all $k$. Since Eq.~(\ref{eq:modeequation}) is a second order equation and the frequency depends on time, the normalization condition is in general not sufficient to specify the positive-frequency modes uniquely, contrary to the case of constant frequency $\omega_0$ for which $h_k^0(\eta) = e^{-i\omega_0 \eta}/(2\omega_0)^{1/2}$. Different boundary conditions for the solutions $h_k(\eta)$ define in general different creation and annihilation operators $a_k$ and $a_k^\dagger$, and thus in general different vacua.\footnote{The precise definition of a vacuum in a curved space-time is still subject to some ambiguities. We refer the interested reader to \protect\citet{Fulling:1979,Fulling:1989,Birrell:1982,Wald:1994} and to the discussion in \protect\citet{Chung:2003} and references therein.}
For example,
solutions which satisfy the condition of having only positive-frequencies in the distant past,
\begin{equation}
h(\eta) \sim e^{-i \omega_k^{-} \eta} \quad {\rm for~}\eta\to -\infty,
\end{equation}
contain both positive and negative frequencies in the distant future,
\begin{equation}
\label{eq:bogolubov}
h(\eta) \sim \alpha_k e^{-i \omega_k^{+} \eta} + \beta_k e^{+ i \omega_k^{+} \eta }  \quad {\rm for~}\eta\to +\infty.
\end{equation}
Here $\omega_k^{\pm} = \lim_{\eta\to\pm\infty} \omega_k(\eta)$. As a consequence, an initial vacuum state is no longer a vacuum state at later times, i.e.\ particles are created. The number density of particles is given in terms of the Bogolubov coefficient $\beta_k$ in Eq.~(\ref{eq:bogolubov}) by
\begin{equation}
n = \frac{1}{(2\pi a)^3} \int d^3k |\beta_k|^2.
\end{equation}

These ideas have been applied to gravitational particle creation at the end of inflation by \citet{Chung:1998} and \citet{Kuzmin:1998}. Particles with masses $M_X$ of the order of the Hubble parameter at the end of inflation, $H_I \approx 10^{-6} M_{\rm Pl} \approx 10^{13} {\rm ~GeV}$, may have been created with a density which today may be comparable to the critical density. Figure 8 shows the relic density $\Omega h^2$ of these WIMPZILLAs as a function of their mass $M_X$ in units of $H_I$. Curves are shown for inflation models that have a smooth transition to a radiation dominated epoch (dashed line) and a matter dominated epoch (solid line). The third curve (dotted line) shows the thermal particle density at temperature $T=H_I/2\pi$. Also shown in the figure is the region where WIMPZILLAs are thermal relics. It is clear that it is possible for dark matter to be in the form of heavy WIMPZILLAs generated gravitationally at the end of inflation. 

\section{Neutralino dark matter searches}

The second part of these lectures is an introduction to several methods to detect non-baryonic dark matter. We will use the lightest neutralino as our guinea pig, because of the variety of techniques that can be employed to detect it, but the discussion is more general and can be applied to a generic WIMP. Thus in this second part we assume that non-baryonic dark matter is made of WIMPs (in particular, neutralinos), and we examine several observational ways to test our assumption.

Neutralino dark matter searches are traditionally divided into two main categories:  (1) direct detection of Galactic dark matter in laboratory experiments, and (2) indirect detection of neutralino annihilation products. For the sake of exposition, indirect searches are further subdivided into: (2a) searches for high-energy neutrinos from the center of the Sun or of the Earth; (2b) searches for anomalous cosmic rays and gamma-rays from galactic halos, especially our own; and (2c) searches for neutrinos, gamma-rays, and radio waves from the Galactic Center. We now examine each of them in turn.

\subsection{Direct detection}

The idea here is that neutralino dark matter is to be found not only in the halo of our galaxy and in our solar system, but also here on Earth and in the room we are in. Thus if we could set up a detector that records the passage of dark matter neutralinos, we could hope of detecting neutralino dark matter. 

A process that can be used for this purpose is the elastic scattering of neutralinos off nuclei. Inelastic scattering could also be used in principle, as could  scattering off electrons, but the rate of these processes are expected to be (much) smaller. 

Dozens of experiments worldwide, too numerous to be all listed here, are using or plan to use elastic scattering to search for neutralino dark matter, or WIMP dark matter in general. The small expected detection rate, and the necessity of suppressing any ionizing radiation passing through the detector, are reasons to shelter these experiments from cosmic rays, e.g.\ by placing them in mines or underground laboratories.

Generally, with the notable exception of directional detectors described below, only the energy deposited in the detector during the elastic scattering can be measured. This energy is of the order of a few keV, for typical neutralino masses and speeds in the galactic halo. The kinetic energy of the recoiling nucleus is converted partly into scintillation light or ionization energy (giving an electric current) and partly into thermal energy (heating up the detector). 

In cryogenic detectors, a simultaneous measurement of both ionization and thermal energy allows the discrimination of nuclear recoils from electrons produced in radioactive decays or otherwise. This discrimination, however, cannot tell if the nuclear recoil was caused by a WIMP or an ambient neutron. The detector, most often a germanium or silicon crystal, needs to be cooled at liquid helium temperature so that its low heat capacity converts a small deposited energy into a large temperature increase. Only relatively small crystals can be currently used in these cryogenic detectors, with relatively low detection rates. 

Detection rates can be increased by using bigger detectors operated at room temperature, at the expense of giving up a measurement of the thermal energy and loosing discrimination power against electrons. The biggest dark matter detector is currently of this type. It is a sodium iodide crystal (a scintillator) under the Gran Sasso mountain in Italy, and it belongs to the Italian-Chinese collaboration DAMA (short for DArk MAtter). Interestingly, the loss of discrimination power and the gain in target mass almost compensate each other, and the sensitivity of cryogenic and scintillation detectors is not very different.

\begin{figure}[!t]
\centering
\includegraphics[width=0.7\textwidth]{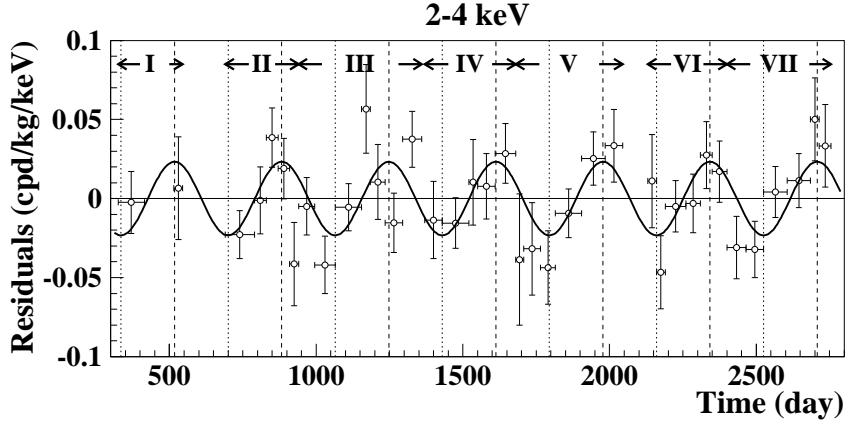}
\vspace{-\baselineskip}
\caption{Annual modulation of the total counting rate (background plus possible dark matter signal) in seven years of data with the DAMA-NaI detector. A constant counting rate has been subtracted to give the `residuals.' The significance of the modulation is 6$\sigma$ and its period is 1 year. The interpretation of the yearly modulation as due to a WIMP signal is controversial. (Figure from \citealp{Bernabei:2003}.) }
\end{figure}

\begin{figure}[!b]
\begin{minipage}{0.39\textwidth}
\caption{Sketch illustrating the directions of the Sun's and the Earth's motions during a year. As the Sun moves in the Galaxy (here at 232 km/s, $60^\circ$ out of the plane of the Earth's orbit), the Earth moves around the Sun (here at 30 km/s). 
The vectorial sum of their velocities gives the velocity of the Earth with respect to the Galaxy. Assuming the WIMPs to be on average at rest in the Galaxy, it follows that the average speed of the WIMPs relative to the Earth is modulated with a period of 1 year.}
\end{minipage}
\begin{minipage}{0.6\textwidth}
\includegraphics[width=\textwidth]{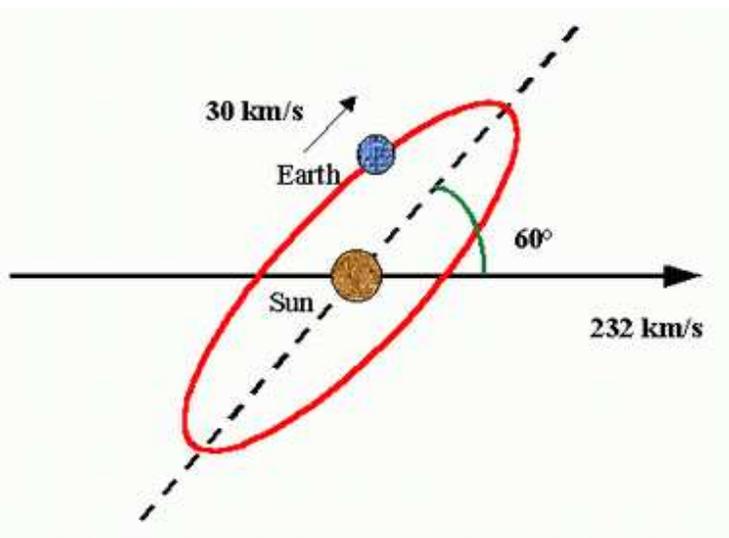}
\end{minipage}
\end{figure}

\subsubsection*{Annual modulation}

Few years ago, the DAMA collaboration reported a possible detection of WIMP dark matter \citep{Belli:1997,Bernabei:1998,Bernabei:1999,Bernabei:2000,Bernabei:2003}. Their most recent data \citep{Bernabei:2003} span 7 years and show a $6.3\sigma$ modulation in their total counting rate (signal+background) with a period of 1 year and an amplitude of $\sim 0.02$ events/day per kg of detector and keV of visible recoil energy (Figure 9). This kind of yearly modulation in a WIMP signal was predicted by \citet{Drukier:1986} and \citet{Freese:1988} on the basis that the velocity of the Earth around the Sun adds vectorially to the velocity of the Sun in the Galaxy to produce a yearly modulation in the average speed of the WIMPs relative to the Earth (the WIMPs are assumed on average at rest in the Galaxy). For an observer on the Earth, the WIMP `wind' arrives at a higher speed when the Earth and the Sun move in the same direction and at a lower speed when they move in opposite directions. (The two velocities are actually misaligned by $\sim 60^\circ$; see Figure 10.) The WIMP flux, and the WIMP detection rate, both  proportional to the relative speed of the Earth and the WIMPs, are similarly modulated. 

While the presence of a yearly modulation in the DAMA data seems now to be established, its interpretation as due to WIMPs is controversial. Firstly, it is not hard to imagine that the background itself can undergo seasonal variations with a period of a year. The DAMA collaboration has examined many possible sources of background variation, including a yearly modulation of the cosmic ray intensity underground due to winter-summer temperature changes in the upper atmosphere. They claim to have found no annual variation in the background level that would produce an amplitude as big as the observed one. Secondly, the EDELWEISS cryogenic detector has sensitivity comparable to DAMA's but has not recorded any nuclear recoil event \citep{Benoit:2002}. And the CDMS-I cryogenic detector, also of comparable sensitivity, has detected nuclear recoil events attributed to ambient neutrons in the shallow site where the detector was running \citep{Akerib:2003}. Comparison of these experimental results is however not as straightforward as it may seem, because the relationship between detection rates in cryogenic and scintillation detectors depends, among other things, on the kind of WIMP-nucleus interaction and on details of the WIMP velocity distribution in the halo, which are both poorly known.

This dependence is apparent in the expression for the expected counting rate per recoil energy bin and unit detector mass $dR/dE$. We have
\begin{equation}
\frac{dR}{dE}  =  \int \frac{N_T}{M_T} \times \frac{ d\sigma}{dE} \times n v f(\vec{v},t) d^3v ,
\end{equation}
where $N_T$ and $M_T$ are the number of target nuclei and the detector mass, respectively, $d\sigma/dE$ is the WIMP-nucleus differential cross section, and $n v f(\vec{v},t)$ is the WIMP flux impinging on the detector. Here $n$ denotes the WIMP density, $v$ the WIMP speed, and $f(\vec{v},t) d^3v$ the WIMP velocity distribution. We write $M_T/N_T=M$, the nuclear mass, $n=\rho/m$, and $d\sigma/dE = \sigma_0 |F(q)|^2/E_{\rm max}$, where $\sigma_0$ is the total scattering cross section of a WIMP off a fictitious point-like nucleus, $|F(q)|^2$ is a nuclear form factor that depends on the momentum transfer $q=\sqrt{2ME}$ and is normalized as $F(0)=1$, and $E_{\rm max}=2\mu^2 v^2/M$ is the maximum recoil energy imparted by a WIMP of speed $v$ ($\mu=mM/(m+M)$ is the WIMP-nucleus reduced mass). Hence
\begin{equation}
\frac{dR}{dE} = \frac{ \rho \sigma_0 |F(q)|^2}{2 m \mu^2} \int_{v>\sqrt{ME/2\mu^2}} \frac{f(\vec{v},t)}{v} d^3 v . 
\end{equation}

Notice that one can only measure the product $\rho \sigma_0$ with this technique.
Notice also that the event rate at energy $E$ depends on the WIMP velocity distribution at speeds $v>\sqrt{ME/2\mu^2}$. This integration limit depends on the nuclear mass, and thus detectors with different kinds of nuclei are sensitive to different regions of the WIMP velocity space. Moreover, the cross section $\sigma_0$ scales differently for spin-dependent and spin-independent WIMP-nucleus interactions. Finally, while there is a consensus on the spin-independent nuclear form factors, spin-dependent form factors are sensitive to detailed modeling of the proton and neutron wave functions inside the nucleus \citep[see][and references therein]{Jungman:1996}. 

For spin-independent interactions with a nucleus with $Z$ protons and $A-Z$ neutrons, one has
\begin{equation}
\label{eq:spinindependent}
  \sigma_0 = \frac{\mu^2}{\pi} \big| Z G^p_s + (A-Z) G^n_s \big|^2 \simeq A^2 \frac{\mu^2}{\pi} \big| G^p_s \big|^2,
\end{equation}
where $G^p_s$ and $G^n_s$ are the scalar four-fermion couplings of the WIMP
with point-like protons and neutrons, respectively \citep[see, e.g,][]{lesarcs}. The last approximation holds for the case $G^p_s \simeq G^n_s$, which is typical of a neutralino. The spin-independent event rate, proportional to $\sigma_0/\mu^2$, scales with the square of the atomic number $A$ (if we neglect the form factor). It is this dependence on $A$ that allows detectors with relatively heavy nuclei to reach down to WIMP-proton cross sections typical of weak interactions. 

For spin-dependent interactions, one has instead
\begin{equation}
  \sigma_0 = \frac{4\mu^2}{\pi} \frac{J+1}{J} \big| \langle S_p \rangle G^p_a + \langle S_n \rangle G^n_a \big|^2 ,
  \end{equation}
  where $J$ is the nuclear spin, $\langle S_p \rangle$ and $ \langle S_n \rangle$ are the expectation values of the spin of the protons and neutrons in the nucleus, respectively, and $G^p_a$ and $G^n_a$ are the axial four-fermion couplings of the WIMP
with point-like protons and neutrons \citep[see][]{lesarcs,tovey}. There is no increase of the spin-dependent rate with $A^2$, and spin-dependent cross sections of the order of weak cross sections are hard to reach with current detector technology.

Given all these ambiguities in the comparison of cryogenic and scintillator results, it is the author's opinion that the important issue if WIMP dark matter has been detected is not settled yet. A bigger DAMA detector and an upgraded CDMS detector running in the low-background Soudan mine are currently taking data. EDELWEISS is also improving their sensitivity, and new experiments, like CRESST-II and ZEPLIN-IV, should start taking data shortly. 

\subsubsection*{Current bounds and future reach}

The sensitivity of some future experiments is shown in Figure 11, together with the current best bounds from the cryogenic detectors CDMS-I \citep{Akerib:2003} and EDELWEISS \citep{Benoit:2002}, and the region where DAMA claims evidence for a WIMP signal \citep{Bernabei:2003}. As it is conventional in comparing results from different experiments, the figure shows the WIMP-proton spin-independent cross section obtained from experimental data using Eq.~(\ref{eq:spinindependent}) under the assumption of a Maxwellian distribution with conventional parameters for the WIMP velocity. For an historical perspective, the figure also displays the first observational bound on WIMP dark matter obtained by \citet{Ahlen:1986}. 

\begin{figure}[t]
\begin{minipage}{0.6\textwidth}
\includegraphics[width=\textwidth]{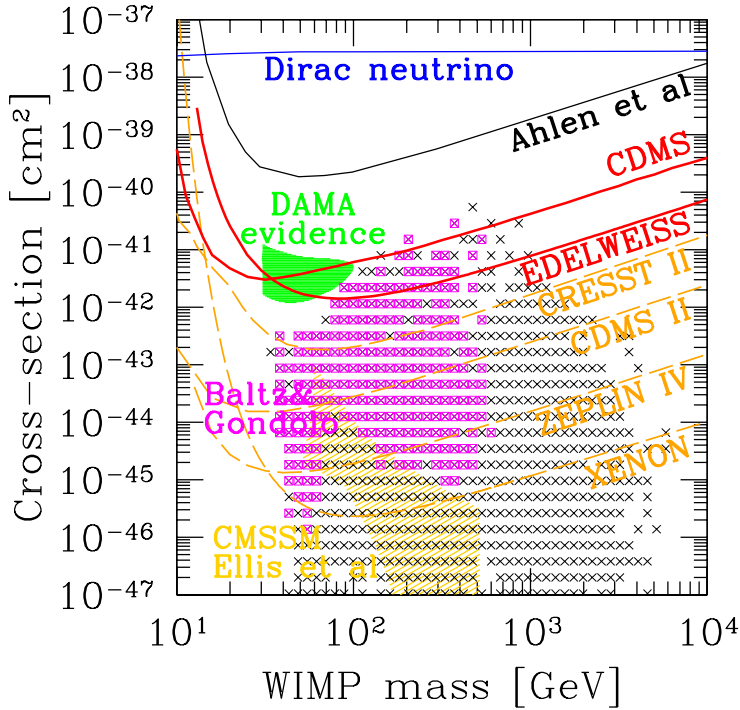}
\end{minipage}
\begin{minipage}{0.39\textwidth}
\caption{Current best bounds (lines labeled `CDMS' and `EDELWEISS') on the dark matter WIMP-proton spin-independent cross section as a function of WIMP mass, together with the expected reach of a few of the upcoming experiments (`CRESST-II', `CDMS-II', `ZEPLIN-IV', `XENON'), some theoretical expectations (`Dirac neutrino', `Baltz \& Gondolo', `CMSSM Ellis et al.'), and the first bound on WIMP dark matter (`Ahlen et al'). See text for details.}
\end{minipage}
\end{figure}

We also indicate theoretical predictions  for a Dirac neutrino with standard model couplings, and for the lightest neutralino in two supersymmetric scenarios: minimal supergravity as in \citet{Ellis:2000} (shaded yellow region) and weak-scale MSSM as in \citet{Baltz:2001,Baltz:2003} (black crosses and magenta squares). Theory models assume that the respective dark matter particles fill up the galactic halo. 
Dirac neutrinos are excluded as main constituents of galactic dark matter in the mass range 3 GeV -- 3 PeV (the bounds continue linearly to the right of the figure). These are the bounds from dark matter searches used in Section 2.2 (and Figure 3) to conclude that a yet-undetected particle species is needed to provide cold dark matter.
For the neutralino, the expected scattering cross section varies in a wide range. Even with the most restrictive assumptions of the Constrained Minimal Supersymmetric Standard Model (CMSSM) in \citet{Ellis:2000}, the cross section at a given neutralino mass can change by an order of magnitude when the other supersymmetric parameters are changed (shaded yellow region in Figure 11). For neutralino masses $m_\chi \sim 400 $ GeV, the CMSSM model parameters may conspire to make the spin-independent cross section arbitrarily small (if such is the case, the total cross section will be dominated by spin-dependent terms, which are however much smaller for the heavy nuclei in current detectors). The highest CMSSM cross sections are within reach of the upcoming experiments, although outside the current best bounds.
Other studies of supergravity models \citep[e.g.][]{Feng:2000} are less restrictive and result in somewhat larger cross sections.
Enlarging the parameter space beyond supergravity, as for example in \citet{Baltz:2001,Baltz:2003},  opens up more possibilities for the values of the cross section. This happens basically for two main reasons; (i)  The supergravity relation between the masses of squark and sleptons on one side and Higgs bosons on the other side is removed; the Higgs boson can thus be lighter, and the spin-dependent scattering cross section, which varies essentially with the fourth inverse power of the Higgs mass, can be larger; (ii) It is no longer required that the electroweak symmetry breaking is achieved radiatively, with the consequence that the gaugino and higgsino content of the lightest neutralino can be arbitrary; the scattering cross section is then enhanced because mixed neutralinos couple to nucleons stronger than pure gauginos or pure higgsinos. This explains the larger extent of the region  covered by the \citet{Baltz:2001,Baltz:2003} models in Figure 11. Finally, the region in Figure 11 marked by the magenta squares indicates the subset of the supersymmetric models examined by \citet{Baltz:2001,Baltz:2003} that could be able to quantitatively explain the $\sim3\sigma$ deviation between the measured value and the Standard Model value of the magnetic moment of the muon.\footnote{The size of the deviation has been hard to determine conclusively because of the difficulty of the non-perturbative QCD calculations involved, in particular  their dependence on the data used as input. There is also a story on sign errors in some of the theoretical calculations....}  These models have relatively light masses for supersymmetric partners, and give neutralino-proton cross sections which are relatively large and within the reach of the most ambitious future experiments.

\subsubsection*{Directional detection}

We conclude this section by mentioning the very intriguing possibility of WIMP detectors that are sensitive to the direction of nuclear recoils (directional detectors). 

The advantages of measuring the recoil direction are multiple: a more powerful background discrimination; the detection of the new modulation effects, such as a daily modulation in the arrival direction of WIMPs due to the Earth rotation around its axis;  and the exciting possibility of reconstructing the WIMP velocity distribution in the solar neighborhood. The latter is possible because of a simple relation between the WIMP velocity distribution $f(\vec{v},t)$ and the nuclear recoil rate differential in both energy $E$ and recoil direction $\uvec{q}$ \citep{Gondolo:Radon},
\begin{equation}
\label{eq:dRdEdOmega}
\frac{ dR } {d{E} d{\Omega}} = 
      \frac{ n \sigma_0 |F(q)|^2}{4 \pi \mu^2 } \, 
      \radon{f}(w,\uvec{q})
\end{equation}
where $d\Omega$ is an infinitesimal solid angle around the direction $\uvec{q}$, $w= \sqrt{ {ME}/{2\mu^2} }$, and 
\begin{equation}
\radon{f}(w,\uvec{q}) = \int \delta ( \vec{u} \cdot \uvec{q} - w) f(\vec{v},t) \, d^3 v 
\end{equation}
is the Radon transform of the WIMP velocity distribution. 

A promising development in this direction is the DRIFT detector
\citep{DRIFT}. This detector consists of a negative ion time projection
chamber, the gas in the chamber serving both as WIMP target and as 
ionization medium for observing the nuclear recoil tracks. The direction of the
nuclear recoil is obtained from the geometry and timing of the image of the
recoil track on the chamber end-plates.  A 1 m$^3$ prototype has been
successfully tested, and a 10 m$^3$ detector is under consideration.

Directional detection is particularly powerful for detecting structure in the dark matter velocity space, as discussed in the next section on the Sagittarius stream.

\subsubsection*{Sagittarius stream}

Recent observations
of the stellar component of the Galactic halo show evidence of a
merger history that has not yet become well mixed, and corroborate previous indications that
halos form hierarchically.  In particular, the Sloan Digital Sky Survey \citep{Newberg:2003} and the Two
Micron All Sky Survey \citep{Majewski:2003} have traced the tidal
stream \citep{Ibata:2001} of the Sagittarius (Sgr) dwarf spheroidal galaxy. The Sagittarius dwarf spheroidal galaxy, of roughly
$10^9 M_\odot$, is a satellite of our own much larger Milky Way
Galaxy, located inside the Milky Way, $\sim$12 kpc behind the Galactic
Center and $\sim$12 kpc below the Galactic Plane \citep{Ibata:1997}.
Two
streams of matter are being tidally pulled away from the main body of
the Sgr galaxy and extend outward from it.  These streams, known as the leading and
trailing tidal tails, are made of matter tidally pulled away from the
Sgr galaxy.  It appears that the leading tail is showering matter down
upon the solar neighborhood \citep{Majewski:2003}.  The flow is in the general
direction orthogonal to the Galactic plane and has a speed of roughly
300 km/s. This speed is comparable to
that of the relative speed of the Sun and the WIMPs in the general
dark halo. 

It is natural to expect that dark matter is associated with the detected tidal streams. Hence one can hope to detect the stream in direct detection
experiments. The detectability depends on the density of dark matter in the stream.
The mass-to-light ratio $M/L$ in the stream is unknown, but is
plausibly at least as large as that in the Sgr main body; in fact, the
$M/L$ in the stream may be significantly larger because the dark
matter on the outskirts of the main body would be tidally stripped
before the (more centrally located) stars.  Various
determinations of the $M/L$ for the Sgr main body give values in the
range 25 to 100 (see the discussion in \citealp{Majewski:2003} and references
therein).  \citet{Freese:2003} and \citet{Freese:2004} have estimated the density
of dark matter in the stream, and find it to be in the range 0.3\% to 23\% of the local (smoothed) dark
halo density. This agrees with a previous theoretical study on the tidal
disruption of satellite galaxies falling into the halo of our own Milky Way by \citet{Stiff:2001}. These authors found that, with probability of order 1, 
the Sun should be situated
within a stream of density $\sim 4$\% of the local Galactic halo
density.

The additional flux of WIMPs from the stream shows up as a 0.3--23\%
increase in the rate of nuclear recoils at energies below a
characteristic energy $E_c$, the highest energy that WIMPs in the
stream can impart to a target nucleus.  Hence, there is a step in the
energy recoil spectrum; the count rate in the detector is enhanced at
low energies, but then returns to the normal value (due to Galactic
halo WIMPs) at all energies above the critical energy $E_c$.  This
feature can be observed as a sharp decrease in the count rate above a
characteristic energy that depends on the mass of the target nucleus,
the mass of the WIMP, and the speed of the stream relative to the
detector. Figures 12(a) and (b) show how the recoil spectrum $dR/dE$ is modified by the presence of the stream for a sodium iodide detector (like in DAMA) and a germanium detector (like for CDMS and EDELWEISS). For the sake of illustration, the plots assume a stream density equal to 20\% of the local halo density.

\begin{figure}[!b]
\begin{minipage}{0.49\textwidth}
\centering
\includegraphics[width=0.95\textwidth]{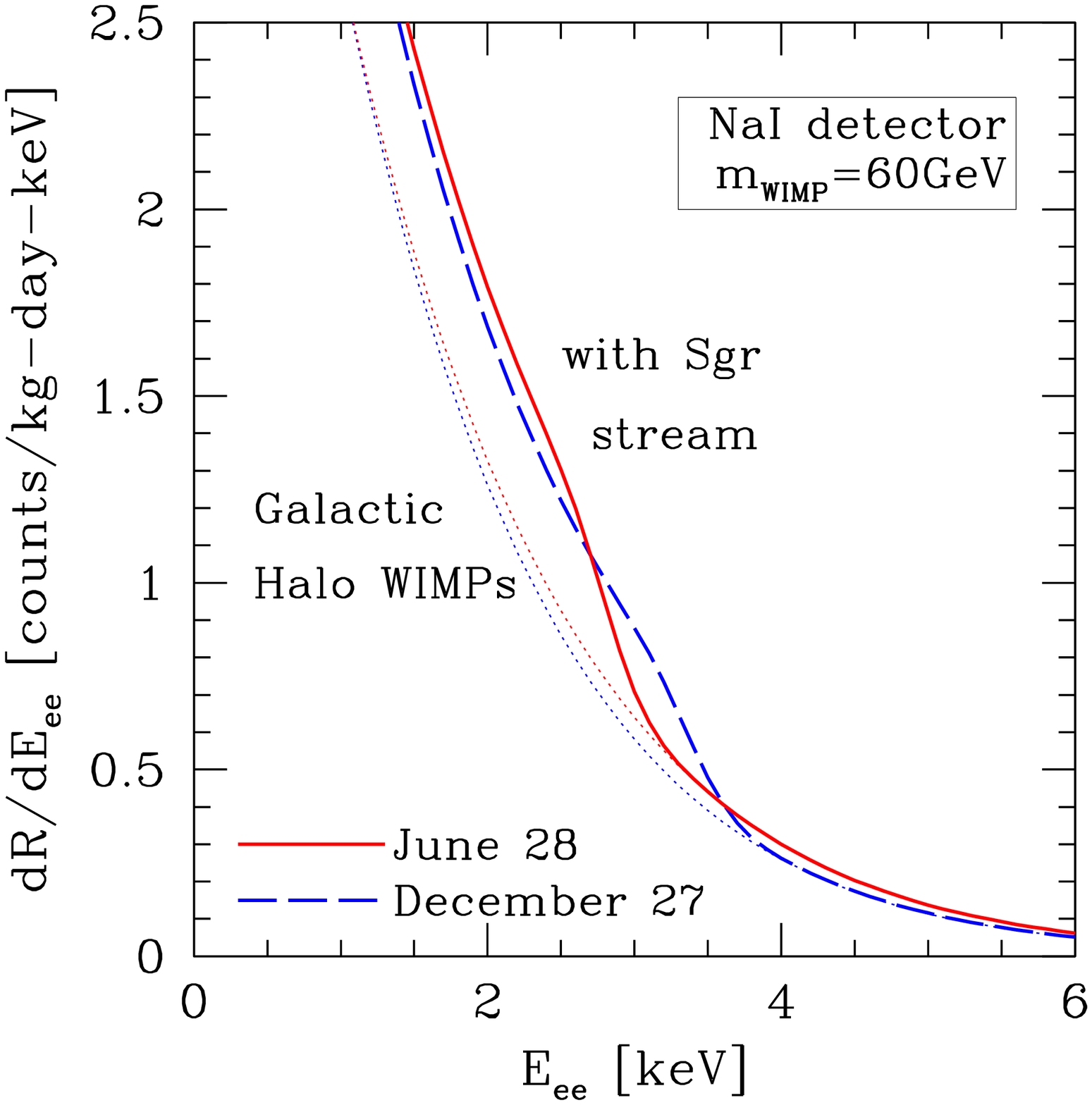}
\end{minipage}
\begin{minipage}{0.49\textwidth}
\centering
\includegraphics[width=0.95\textwidth]{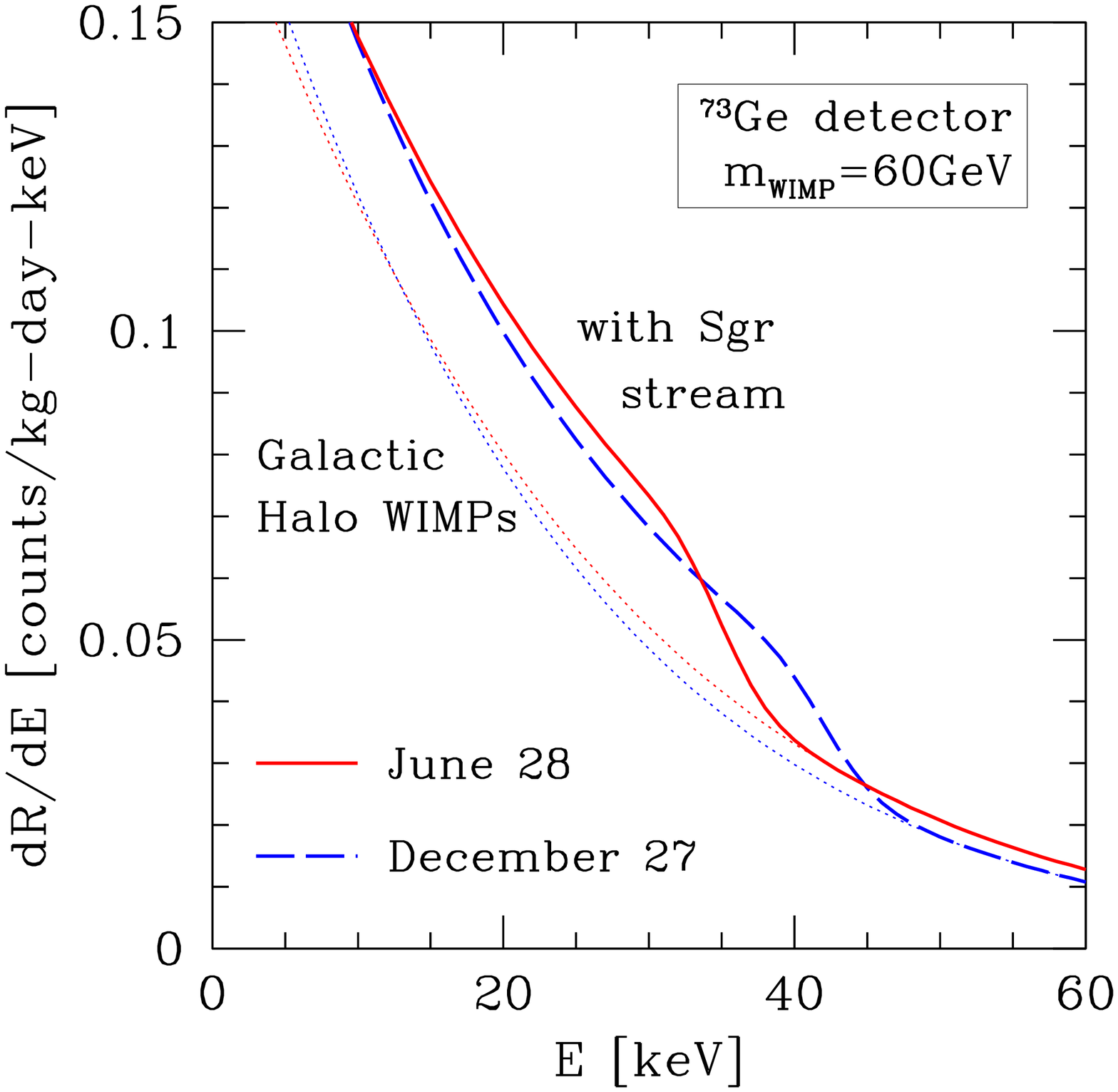}
\end{minipage}
\caption{Effect of the presence of WIMPs in the Sagittarius leading tidal arm.   Count rate of 60 GeV WIMPs in (a) an NaI detector such as DAMA and (b) a Ge detector such as CDMS and EDELWEISS, as a
  function of recoil energy.  The dotted lines (towards the left)
  indicate the count rate due to Galactic halo WIMPs alone for an
  isothermal halo.  The solid and dashed lines indicate the step in
  the count rate that arises if we include the WIMPs in the Sgr stream
  for $v_{str} = 300$ km/s in the direction
  $(l,b)=(90^\circ,-76^\circ)$ with a stream velocity dispersion of 20
  km/sec.  The plot assumes that the Sgr stream contributes an
  additional 20\% of the local Galactic halo density.  The solid and
  dashed lines are for June 28 and December 27 respectively, the dates
  at which the annual modulation of the stream is maximized and
  minimized. (Figure from \citealp{Freese:2003}.)}
\end{figure}

Excitingly, the effect of the stream should be detectable in DAMA, CDMS-II, and other upcoming detectors, and may already be present in the current DAMA data. A detail calculation \citep{Freese:2003} predicts the presence of stream WIMPs in the data with a significance of 100$\sigma$ for DAMA and 11$\sigma$ for CDMS if the stream density is 20\% of the local halo density, and a significance of 24$\sigma$ for DAMA and 3$\sigma$ for CDMS if the stream density is 4\% of the local halo density. (These significance figures are however very sensitive to the velocity assumed for the stream, cfr.~\citealp{Freese:2004}.) 

\begin{figure}[!t]
\begin{minipage}{0.49\textwidth}
\centering
\includegraphics[width=0.95\textwidth]{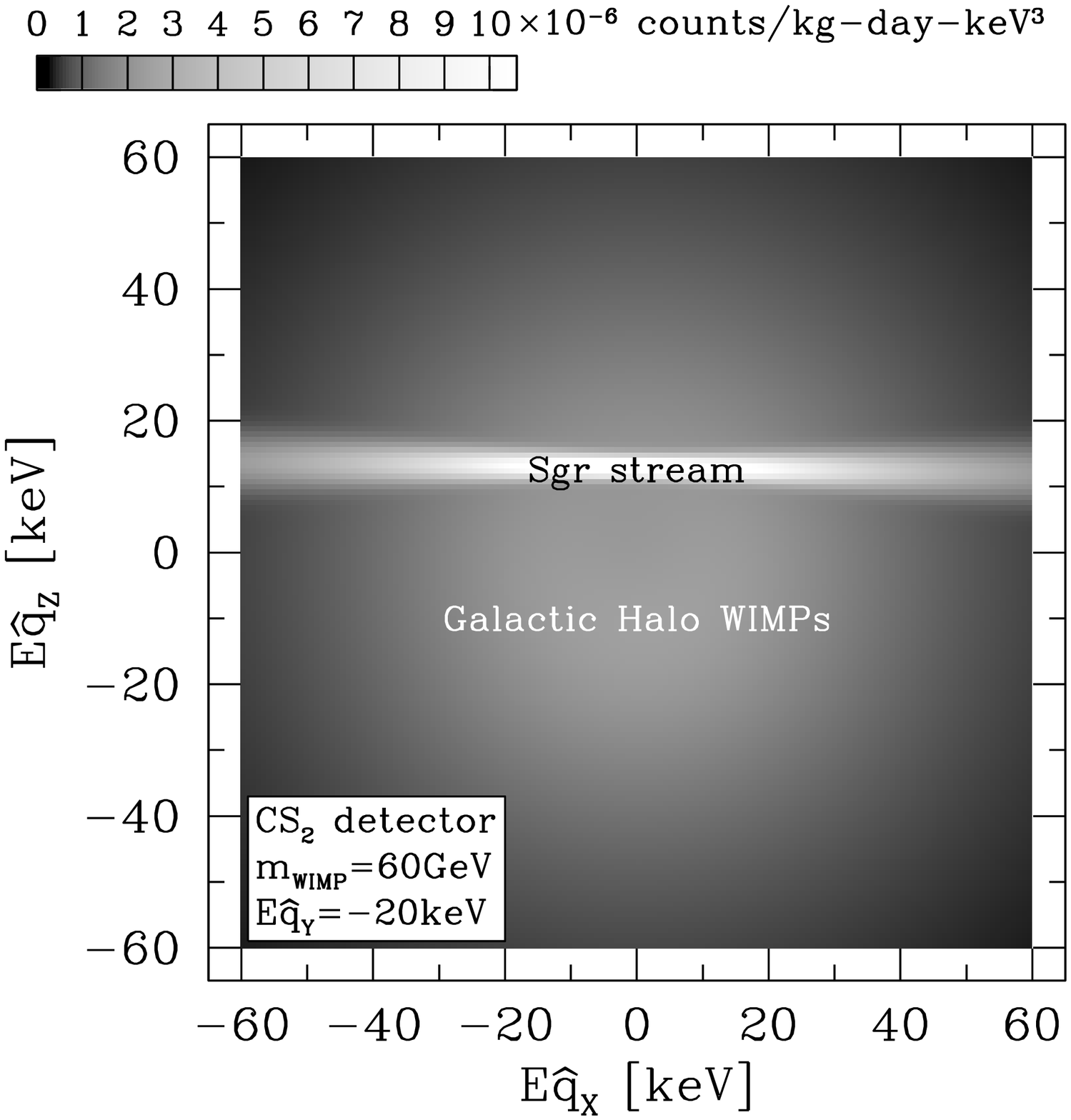}
\end{minipage}
\begin{minipage}{0.49\textwidth}
\centering
\includegraphics[width=0.95\textwidth]{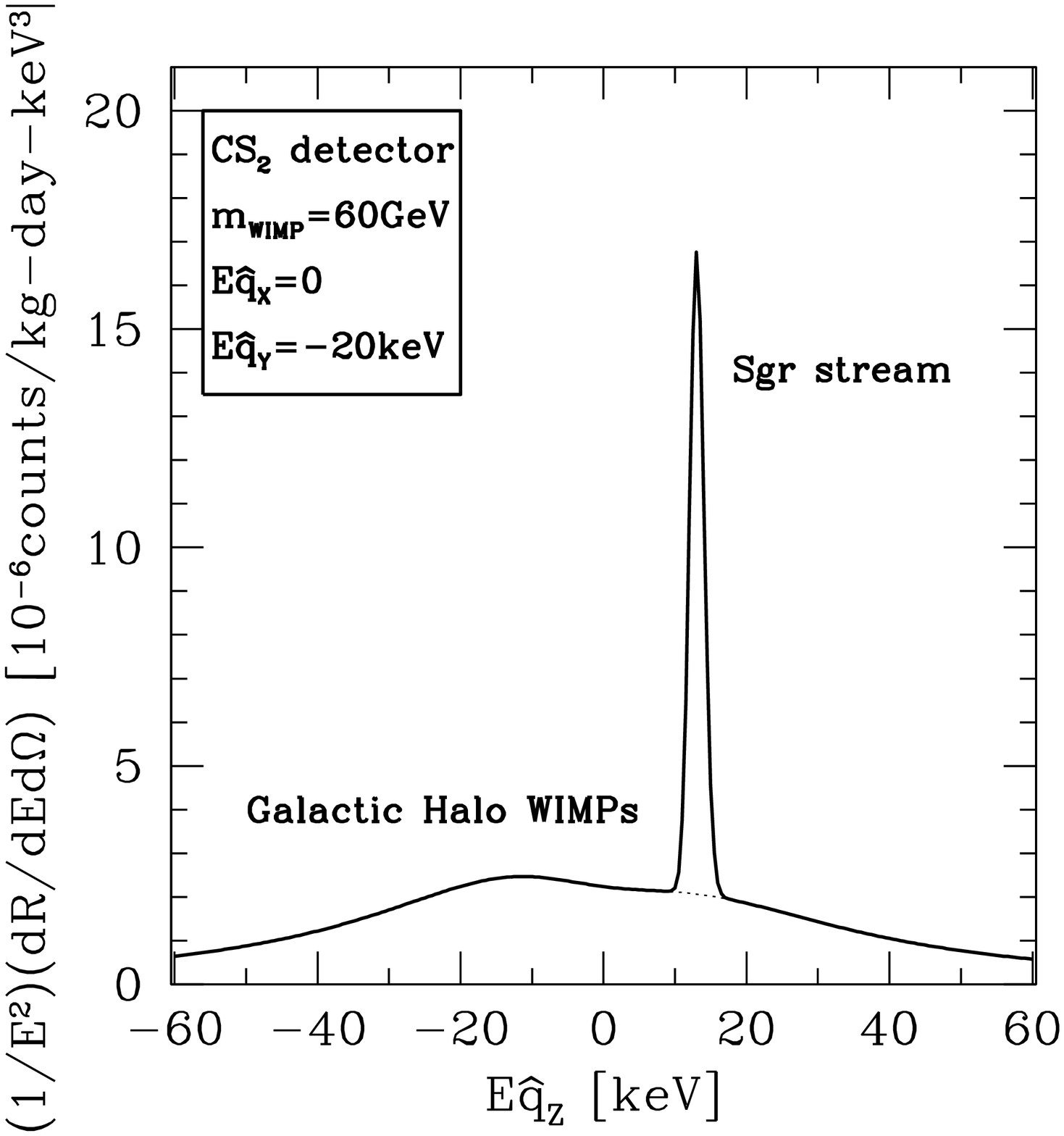}
\end{minipage}
\caption{ Count rate of 60 GeV WIMPs in a CS$_2$ detector (DRIFT) as a
  function of recoil energy $E$ and direction of the nuclear recoil $(\uvec{q}_X,\uvec{q}_Y,\uvec{q}_Z)$. Here $X$ points toward the galactic center, $Y$ toward the direction of galactic rotation, and $Z$ toward the North Galactic Pole. On the left is a density plot of the count rate in the 2-dimensional slice $E\uvec{q}_Y=-20{\rm ~keV}$. On the right is the 1-dimensional section through $E\uvec{q}_Y=-20{\rm ~keV}$ and $\uvec{q}_X=0$. The horizontal axis represents recoils in
  the direction of the Galactic center (left) and Galactic anticenter
  (right); the vertical axis represents recoils in the direction of
  the North Galactic Pole (upward) and South Galactic Pole (downward).
  The gray scale indicates the count rate per kilogram of detector per
  day and per unit cell in the 3-dimensional energy space $E\uvec{q}$. Lighter
  regions correspond to higher count rates. The white band in the
  upper part is the location of nuclear recoils due to WIMPs in the Sgr
  stream. The fuzzy gray cloud at the center contains recoils due to
  WIMPs in the local isothermal Galactic halo. The two WIMP
  populations can in principle be easily separated, given a sufficiently long
  exposure. (Figure from \citealp{Freese:2003}.)}
\end{figure}

Directional detection will be a fantastic means of recognizing the presence of a dark stream through the Solar system. The recoil distribution due to WIMPs in a stream is very much different from the recoil distribution due a Maxwellian velocity distribution. The corresponding Radon transforms that appear in Eq.~(\ref{eq:dRdEdOmega}) are: for a stream of velocity $\vec{V}$,
\begin{equation}
\radon{f}(w,\uvec{q}) \simeq \delta( w - \vec{V} \cdot \uvec{q} )
\end{equation}
which is non-zero only on the surface of a sphere in $(w \uvec{q}_x,w \uvec{q}_y, w \uvec{q}_z)$ space; for a Maxwellian of bulk velocity $\vec{V}$ and velocity dispersion $\sigma$,
\begin{equation}
\radon{f}(w,\uvec{q}) = \frac{ 1}{\sqrt{2\pi\sigma^2}} \exp\left[ - \frac{ ( w - \vec{V} \cdot \uvec{q} )^2} {2 \sigma^2} \right] ,
\end{equation}
which is a smooth gaussian distribution. For our Sgr stream, consider a next-generation DRIFT detector of 30 m$^3$ (DRIFT-2). The difference in the recoil direction distributions of stream and isothermal WIMPs is apparent in Figure 13, where we plot the differential detection rate $E^{-2} dR/dEd\Omega$ for DRIFT-2 under the assumption of a 20\% stream density. Aa seen in the figure, a large DRIFT detector will have the capability of clearly identifying WIMPs in the Sgr stream.

\subsection{Indirect detection}

Besides the direct detection of galactic neutralino dark matter in the laboratory, we can search for dark matter neutralinos by looking for the products of their annihilation. We distinguish three types of searches according to the place where neutralino annihilations occur. The first is the case of neutralino annihilation in the Sun or the Earth, which gives rise to a signal in high-energy neutrinos; the second is the case of neutralino annihilation in the galactic halo, or in the halo of external galaxies, which generates gamma-rays and other cosmic rays such as positrons and antiprotons; the third is the case of neutralino annihilations around black holes, in particular around the black hole at our Galactic Center.

All these annihilation signals share the property of being proportional to the {\it square} of the neutralino density. This follows from the fact that the neutralino is a Majorana fermion, i.e.\ is identical to its antiparticle. Two neutralinos can annihilate to produce standard model particles. Simple stoichiometry then tells us that the annihilation rate, being proportional to the product of the densities of the initial particles, is proportional to the square of the neutralino density. In more detail, we have
\begin{equation}
\label{eq:gamma_ann}
\Gamma_{\rm ann} = \frac{ \sigma_{\rm ann} v \rho^2}{m^2} ,
\end{equation}
where $ \Gamma_{\rm ann} $ is the neutralino annihilation rate per unit volume (i.e.\ the number of neutralinos that are annihilated per unit volume and unit time), $ \sigma_{\rm ann} $ is the neutralino-neutralino annihilation cross section, $ v $ is the relative speed of the two annihilating neutralinos, $\rho$ is the neutralino mass density, and $ m $ is the neutralino mass. Recall that the annihilation cross section $\sigma_{\rm ann}$ goes as $1/v$ at small speeds, as required by kinematical arguments, and thus the product $\sigma_{\rm ann} v $ does not vanish linearly with $v$ (and is not small at the relatively small speeds of neutralinos in galactic halos).
Notice also that the number of annihilations per unit volume and unit time is given by $\tfrac{1}{2} \Gamma_{\rm ann}$, where the factor of $\tfrac{1}{2}$ correctly converts between the number of annihilation events and the number of neutralinos that are annihilated (2 per annihilation). It is easy to get confused with this factor of $\tfrac{1}{2}$.

If in the annihilation rate $\Gamma_{\rm ann}$ we insert a typical weak interaction cross section and a typical value for the average dark matter density in the Universe, the annihilation rate we obtain gives undetectably small signals. Indirect detection is possible because dark matter is not distributed uniformly in space. Galaxies and clusters of galaxies are overdensities in the dark matter field, as is any possible substructure in galactic halos. Furthermore, dark matter may be concentrated gravitationally around massive objects, and may even get trapped inside planets and stars. The neutralino annihilation rate, proportional to the square of the neutralino density, increases substantially in these dark matter concentrations, sometimes to the point of giving observable signals.

\subsubsection{High energy neutrinos from the core of the Sun or of the Earth}

\begin{figure}[!t]
\centering
\includegraphics[width=0.8\textwidth]{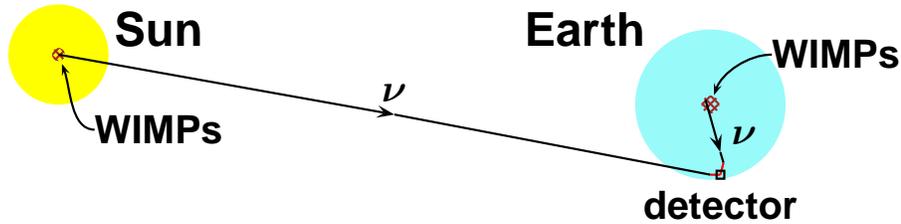}
\caption{Illustration of indirect detection of WIMPs using high-energy neutrinos emitted in WIMP annihilations in the core of the Sun or of the Earth.}
\end{figure}

\begin{figure}[!b]
\begin{minipage}{0.39\textwidth}
\caption{Principle of operation of a water Cherenkov neutrino telescope. An incoming muon neutrino (here a muon neutrino) is converted into a charged lepton (a muon) in the material surrounding the detector (in the rock at the bottom of the sea). The charged lepton moves faster than light in water and thus Cherenkov light (blue cone) is emitted along its trajectory. The Cherenkov light is collected by the array of photomultipliers suspended on strings. Cherenkov neutrino telescopes in ice work on the same principle. (Background figure by Fran\c cois Montanet, ANTARES Collaboration; tracks and Cherenkov light by the present author.)}
\end{minipage}
\hfill
\begin{minipage}{0.5\textwidth} 
\includegraphics[width=0.85\textwidth]{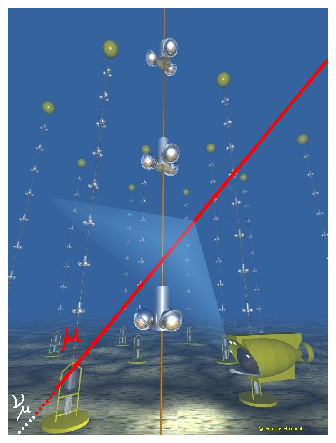}
\end{minipage}
\end{figure}

Neutralinos floating around the solar system can occasionally collide with nuclei in the Sun and in the planets (the Earth, in particular). In these collisions, they may loose enough kinetic energy to end up with a speed smaller than the escape speed, thus becoming gravitationally trapped. After some time, the trapped neutralinos will sink to the core of the celestial body in which they are captured, and will possibly reach a condition of thermal equilibrium (Figure 14). 

Once concentrated in the center, neutralinos annihilate copiously. The annihilation rate is maximal when all captured neutralinos annihilate (a condition called equilibrium between capture and annihilation). Whether this condition is satisfied depends on the relative strength of the annihilation and scattering cross sections, and ultimately on the parameters of the particle and halo models. (See \citealp{Jungman:1996} and references therein for complete formulas.) 

Of the annihilation products produced in the center of the Earth and the Sun only the neutrinos make it to the surface; all the other products are absorbed or decay within a short distance of production. All three flavors of neutrinos are produced for neutralino masses which are currently allowed. Direct production of a neutrino pair is however strongly suppressed in neutralino annihilation, due to the Majorana nature of the neutralino. Annihilation neutrinos are instead produced as secondaries in the decay chains of the primary particles produced in the neutralino-neutralino annihilation. As a consequence, the neutrino energy spectrum is a continuum, and the typical energy of neutrinos from neutralino annihilations is about a tenth of the neutralino mass. Given the current constraints, this means a neutrino energy between few GeVs and few TeVs.

Neutrinos of this energy can be detected in Cherenkov neutrino telescopes, whose principle of operation is depicted in Figure  15. A charged-current interaction in the material surrounding the detector (rock, ice, water) converts the neutrino into its corresponding charged lepton, which then radiates Cherenkov light in the detector medium (ice or water). Several neutrino telescopes are currently operational (among them the Super-Kamiokande detector in Japan and the AMANDA detector at the South Pole), and others are under construction or development (IceCube at the South Pole, ANTARES and NESTOR in the Mediterranean). Other neutrino telescopes have played a role in dark matter searches in the past, such as the IMB, the Fr\'ejus, the MACRO, and the Baksan experiments.

\begin{figure}[!t]
\begin{minipage}{0.49\textwidth}
\includegraphics[width=\textwidth]{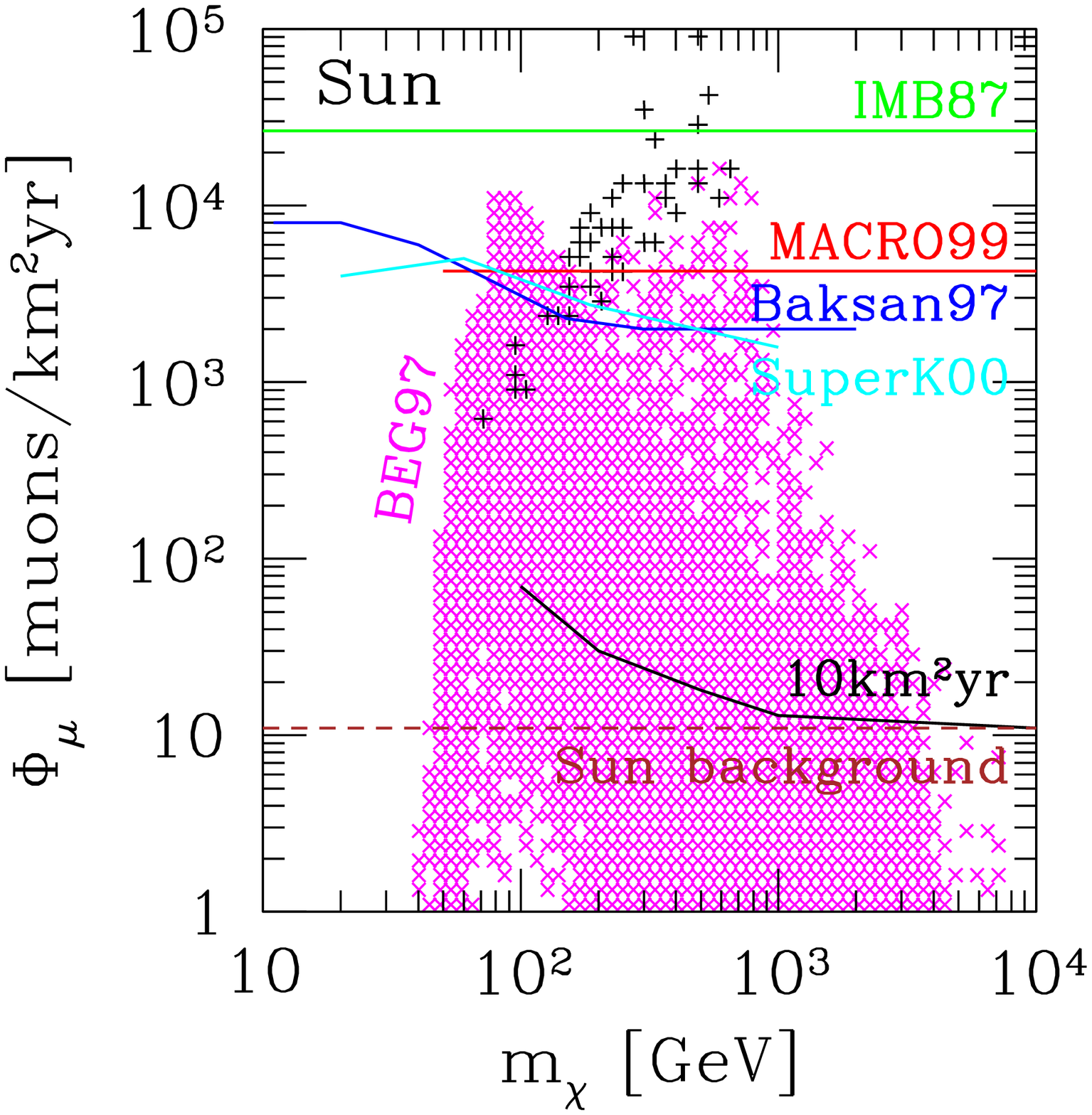}
\end{minipage}
\begin{minipage}{0.49\textwidth}
\hskip0.1\textwidth
\includegraphics[width=0.9\textwidth]{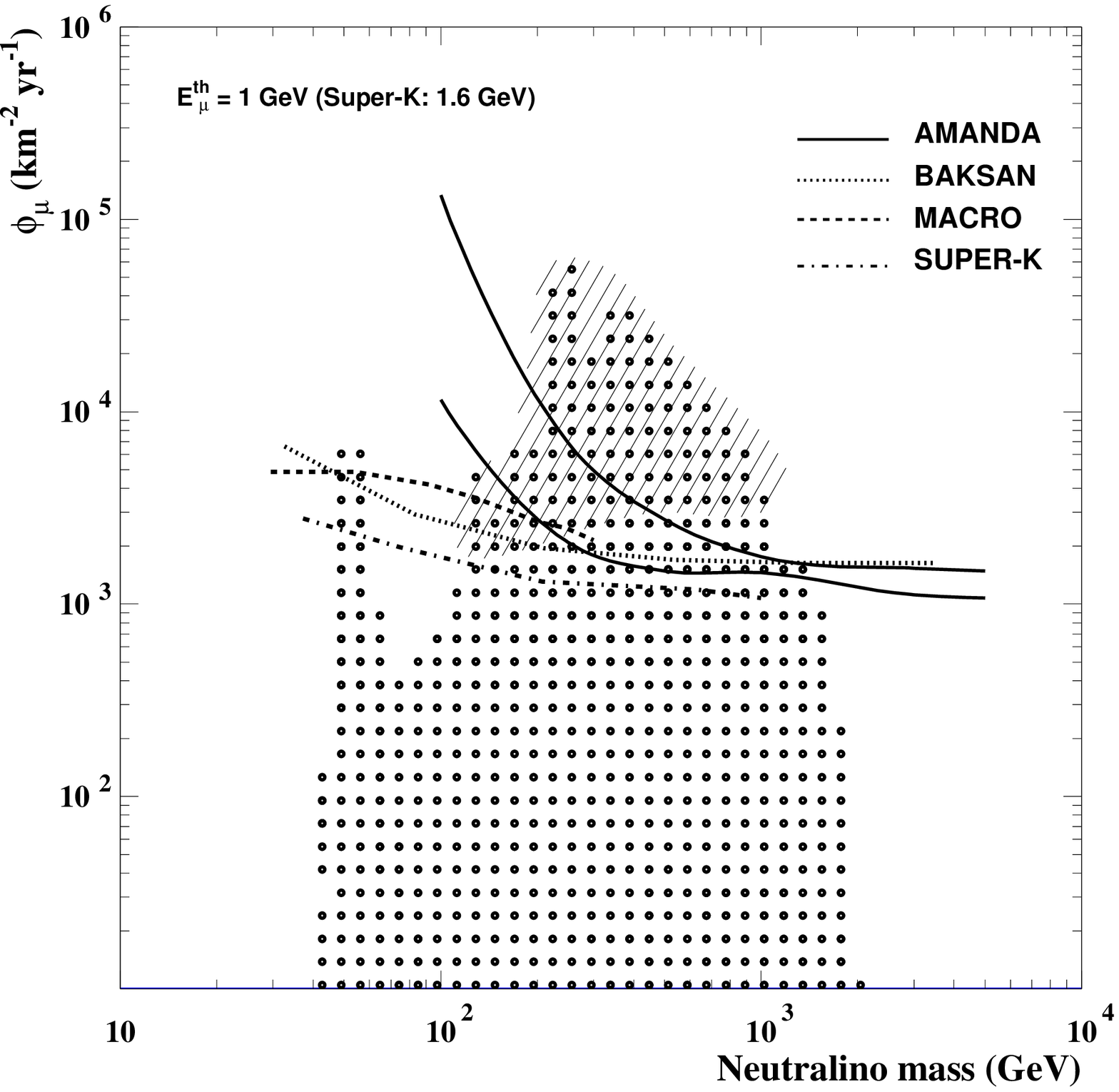}
\end{minipage}
\caption{Indirect searches for neutralino dark matter using high-energy neutrinos from (left) the Sun and (right) the Earth. On the vertical axis is the flux of neutrino-induced muons that traverse the neutrino telescope, on the horizontal axis is the neutralino mass. `IMB87' is the historically first upper limit, `MACRO,' `BAKSAN,' `SuperK,' and `AMANDA' are the limits from the corresponding experiments. The regions marked by $\times$ on the left and by dots on the right are the predictions of supersymmetric models defined at the weak scale \citep{Bergstrom:1997,AMANDA:2002}. The + signs on the left and the shaded region on the right indicate the regions where there are models that  have been excluded by direct dark matter searches. The line labeled `10km$^2$yr' shows the maximum reach of such an exposure in IceCube, and the line labeled `Sun background' marks the level of high-energy neutrino emission due to cosmic ray interactions on the surface of the Sun, which is the ultimate applicability limit  of this method.
 (Figure on the left from \citealp{Gondolo:2000}; figure on the right from \citealp{AMANDA:2002}.) }
\end{figure}

The current experimental situation for this indirect detection method is summarized in Figure 16(a) for neutrinos from WIMPs in the Sun, and Figure 16(b) for neutrinos from WIMPs in the Earth. The figures show the current best bounds from the MACRO, Baksan, Super-Kamiokande, and AMANDA experiments, as well as the first bound obtained using this technique by the IMB collaboration in 1987. Also shown is the reach of the IceCube experiment after an exposure of 10 km$^2$ yr, and the ultimate applicability of this method for the Sun, which is set by the emission of high-energy neutrinos in cosmic ray interactions with nuclei on the surface of the Sun. 

Expectations from theoretical models in Figure 16 (MSSM with seven weak-scale parameters) range from cases which are already excluded by this method to cases which this indirect method will be unable to explore. In comparison, direct searches have a different coverage of theoretical models.
The reach of direct searches is indicated in the Figure 16 by + signs in the left panel and by the shading in the right panel (direct searches exclude only some of the theoretical models that are projected onto these regions from the higher-dimensional supersymmetric parameter space). There are models that can be explored by direct searches and not by indirect searches of high-energy neutrinos from the Sun and the Earth. And vice versa, there are theoretical models that cannot be explored by direct searches but can be explored by indirect searches of high-energy neutrinos from the Sun or the Earth. The latter aspect  is illustrated in Figure 17, where models that can be reached by indirect searches for neutrinos from neutralinos in the Sun are marked by dots in the scattering cross section--mass plane, and compared with the sensitivity of several current and future direct search experiments. Several models fall beyond the reach of even the most ambitious direct dark matter searches. This shows the (everlasting) complementarity between direct and indirect neutralino searches. 

\begin{figure}[!b]
\begin{minipage}{0.34\textwidth}
\caption{Complementarity of direct and indirect neutralino dark matter searches. The figure shows that several supersymmetric models that are within the (expected) reach of a big neutrino telescope of 10 km$^2$ yr exposure are beyond the reach of current and future direct detection experiments. The vertical axis is the product of the neutralino-proton spin-independent scattering cross section $\sigma_{\rm \chi-p}$ and the local neutralino density in units of 0.3 GeV/cm$^3$ $f_{CCDM}$. The horizontal axis is the neutralino mass $m_{\chi}$. (Figure from \citealp{Duda:2002}.)}
\end{minipage}
\hfill
\begin{minipage}{0.66\textwidth}
\centering\includegraphics[width=0.9\textwidth]{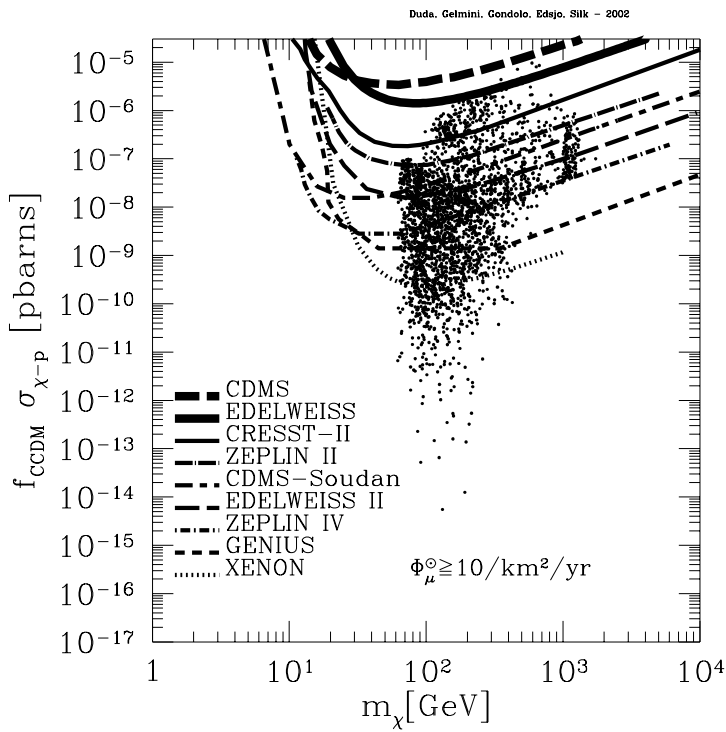}
\end{minipage}
\end{figure}

\subsubsection{Gamma-rays and cosmic rays from neutralino annihilation in galactic halos}

We shift now our attention to signals originating in neutralino annihilations which occur in the halo of our galaxy or in the halo of external galaxies. 

The annihilation products of importance are those that are either rarely produced in astrophysical environments or otherwise have a peculiar characteristic that make them easily recognizable. In the first category are rare cosmic rays such as positrons, antiprotons, and antideuterons. In the second category are gamma-rays, whose spectrum is expected to contain a gamma-ray line at an energy corresponding to the neutralino mass (besides a gamma-ray continuum; see Figure 18). The gamma-ray line is produced directly in the primary neutralino annihilation into $\gamma\gamma$ or $Z\gamma$. Positrons, antiprotons, deuterons, and the gamma continuum are generated in the particle cascades that follow the decay of the primary annihilation products. Their spectra are therefore broad, with a typical energy which is only a fraction of the neutralino mass, and a shape whose details depend on which annihilation channels are dominant. Two neutralinos can in fact annihilate into a variety of primary products, depending on their masses and compositions: fermion pairs $f\bar{f}$, Higgs boson pairs $H_i H_j$, gauge boson pairs $W^+W^-$, $ZZ$, etc.

\begin{figure}[t]
\begin{minipage}{0.53\textwidth}
\vskip10pt
\includegraphics[width=\textwidth]{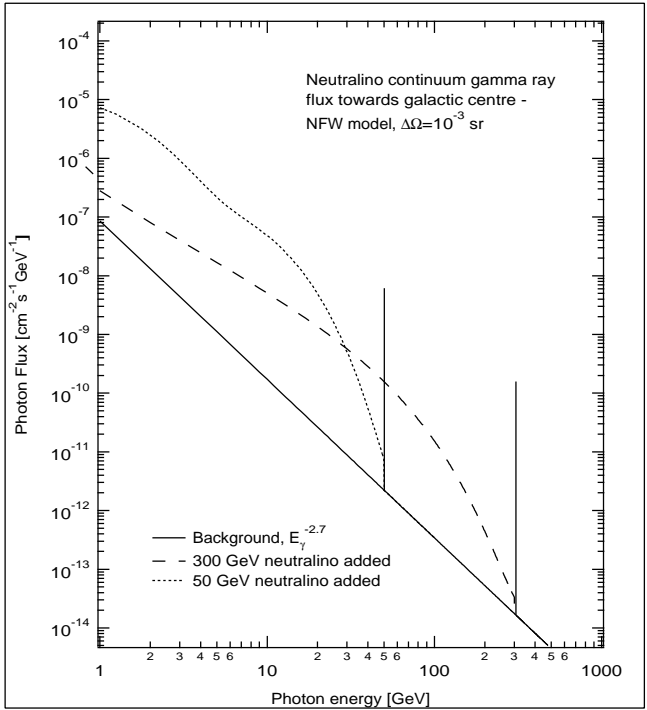}
\end{minipage}
\hfill
\begin{minipage}{0.4\textwidth}
\caption{Two examples of gamma-ray spectrum from neutralino annihilation in the galactic halo. The numerical values refer to a specific model for the dark halo of our galaxy and for observations in the direction of the galactic center, but the spectral characteristics are general. Neutralino annihilations produce a gamma-ray line at an energy corresponding to the mass of the neutralino, and a gamma-ray continuum generated in the particle cascades following the primary annihilation. The figure illustrates that the shape of the continuum spectrum depends on the neutralino mass, but notice that it also depends on the neutralino composition. (Figure from \citealp{BUB:1996}.)}
\end{minipage}
\end{figure}

\begin{figure}[!b]
\begin{minipage}{0.65\textwidth}
\includegraphics[width=0.9\textwidth]{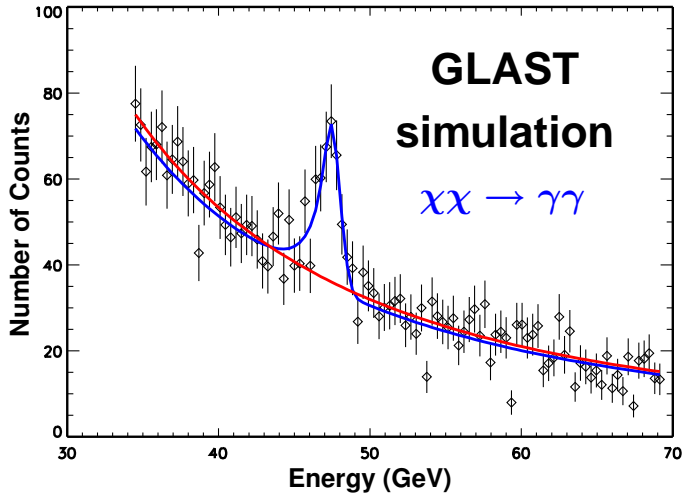}
\vspace{-20pt}
\end{minipage}
\begin{minipage}{0.34\textwidth}
\caption{Simulation of a gamma-ray annihilation line from the annihilation of $\sim$48 GeV neutralinos, superimposed on a gamma-ray background of astrophysical origin. The simulation includes the finite energy resolution of the upcoming GLAST detector. (Figure from the GLAST Science Brochure.)}
\end{minipage}
\end{figure}

Detection of the gamma-ray line would be a smoking-gun for neutralino dark matter, since no other astrophysical process is known to produce gamma-ray lines in the 10 GeV -- 10 TeV energy range. Good energy resolution is crucial to detect the neutralino gamma-ray line. The simulation in Figure 19 shows that the upcoming GLAST detector should have an adequate energy resolution. 

For these signals, the dependence of the annihilation rate on the square of the density, see Eq.~(\ref{eq:gamma_ann}), has dramatic consequences. The predicted signals may change by several orders of magnitude when the model for the dark matter density is changed, even without violating observational bounds on the latter. Truly, these observational limits are not very stringent, given the understandable difficulty of  measuring the dark matter density. Anyhow the problem is currently there, and can be divided roughly into two questions: (1) what is the radial dependence of the average dark matter density in a galaxy, especially in our own? (2) how much substructure, i.e.\ clumps and streams, is there in galactic dark halos?

\subsubsection*{Dependence on the density profile}

Historically, the density profile of galactic dark halos has been given in terms of empirical density profiles whose density is  constant in a central region and decreases as $r^{-2}$ at large radii. The latter is the main ingredient in obtaining a flat rotation curve in the outer regions, which is a primary evidence for dark matter in galaxies. Central among these functions is the cored isothermal profile 
\begin{equation}
\rho_{\rm BS}(r) = \frac{ \rho_0 a^2 }{ r^2+a^2 },
\end{equation}
where $a$ is called the core radius \citep{Bahcall:1980}. This parametrization is so simple and so much used that it is sometimes called the `canonical' density profile. Another interesting empirical parametrization is the density profile of \citet{Persic:1997},
\begin{equation}
\rho_{\rm PS} = \rho_0 \frac{a^2(r^2+3 a^2)}{3(r^2+a^2)^2} ,
\end{equation}
which provides good fits to rotation curves of hundreds of spiral galaxies (which are not as flat as one might think!). 

\begin{figure}[!t]
\begin{minipage}{0.39\textwidth}
\caption{Dark matter density profiles for a galaxy resembling our own. Models `BS' \citep{Bahcall:1980} and `PS' \citep{Persic:1997} are empirical parametrizations which possess a central region with constant density (core). Models `NFW' \citep{Navarro:1996} and `Moore et al' \citep{Moore:1998} are derived from numerical simulations of structure formation in the Universe, and in them the density in the central region increases as a power law of radius (cusp). All four models are normalized to the same total mass and virial radius.}
\end{minipage}
\hfill
\begin{minipage}{0.57\textwidth}
\vskip16pt
\includegraphics[width=0.9\textwidth]{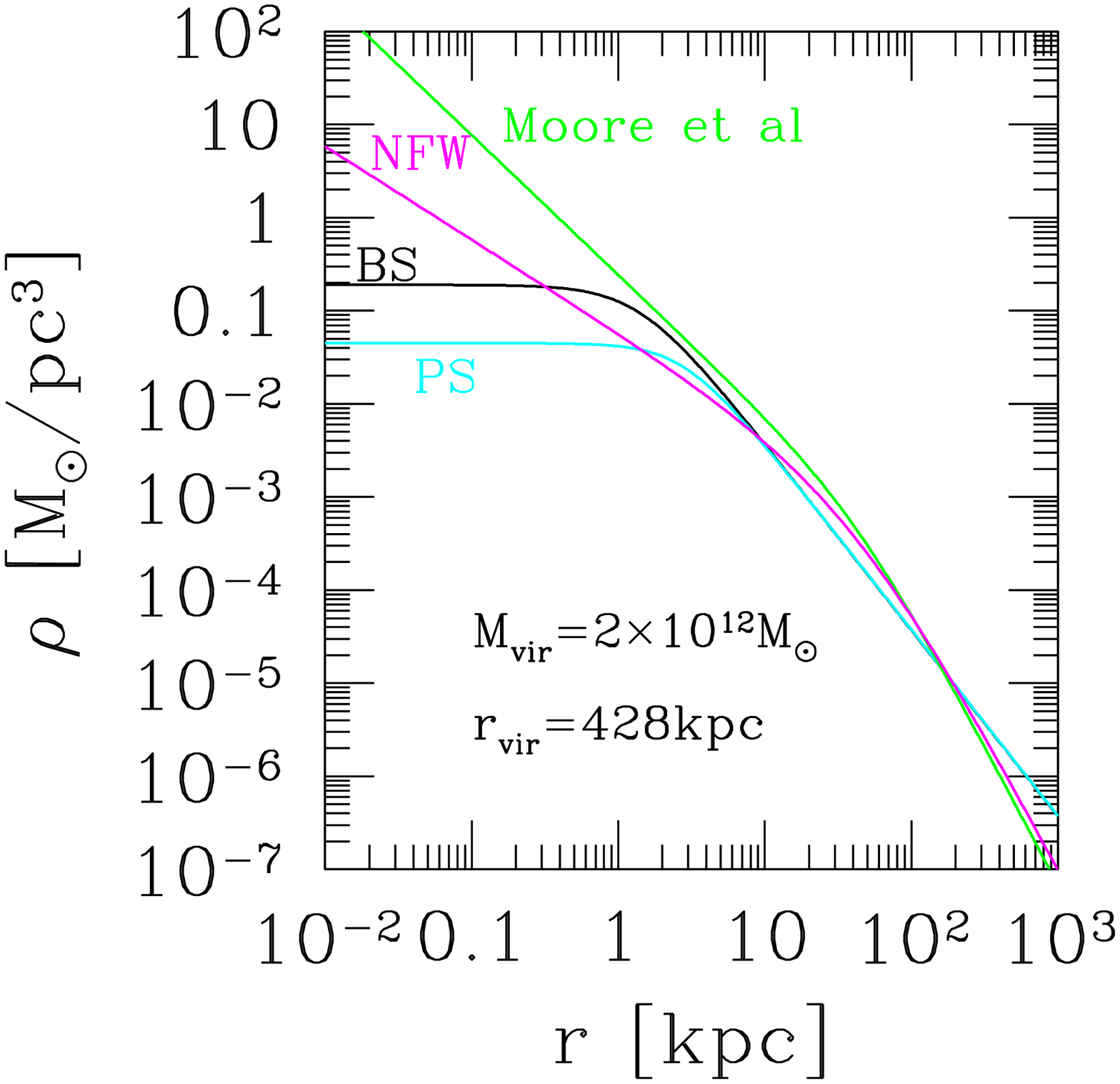}
\end{minipage}
\end{figure}

Numerical simulations of structure formation in the Universe have discovered that pure cold dark matter halos do not follow the previous empirical density profiles but instead have a universal shape whose parameters depend on the mass (or age) of the system. \citet{Navarro:1996} have found this universal profile to have the form 
\begin{equation}
\rho_{\rm NFW}(r) = \frac{ \rho_s r_s^3 } { r (r+r_s)^2 } ,
\end{equation}
where the parameter $r_s$ is the radius at which the radial dependence of the density changes from $r^{-1}$ to $r^{-3}$. The empirical dependence $r^{-2}$ is then seen as an approximation in the transition region. \citet{Moore:1998} suggest instead that the universal profile may be steeper at the center, 
\begin{equation}
\rho_{\rm Moore}(r) = \frac{  \rho_s r_s^3 } { r^{3/2} (r+r_s)^{3/2} }.
\end{equation}
Which of these two profiles better represents the results of numerical simulations is a question that must be answered by higher resolution simulations (which seem to be pointing to an inner slope $\gamma$ that depends on the mass of the system).

The four profiles mentioned above are plotted in Figure 20 for a galaxy that could be our own, with total mass $M_{\rm vir} = 2 \times 10^{12} M_\odot$ and virial radius $ r_{\rm vir} = 428 $ kpc. 

The essential difference between the empirical and the numerical profiles, for what concerns neutralino signals, is their behavior at small radii. The empirical profiles have a central region with constant density, called a core, while the numerical profiles have a density that increases toward the center as a power law, called a cusp. Since the neutralino annihilation signals scale as the square of the density, any power law $r^{-\gamma}$ with $\gamma>0$ is bound to give a higher signal from the central region, for a given total mass. Moreover, if $\gamma\ge 3/2$, the annihilation rate in the central region formally diverges, because $ \int \rho^2 r^2 dr \propto \int r^{-2\gamma+2} dr$.

\begin{figure}[!t]
\label{fig:20}
\begin{minipage}{0.6\textwidth}
\includegraphics[width=\textwidth]{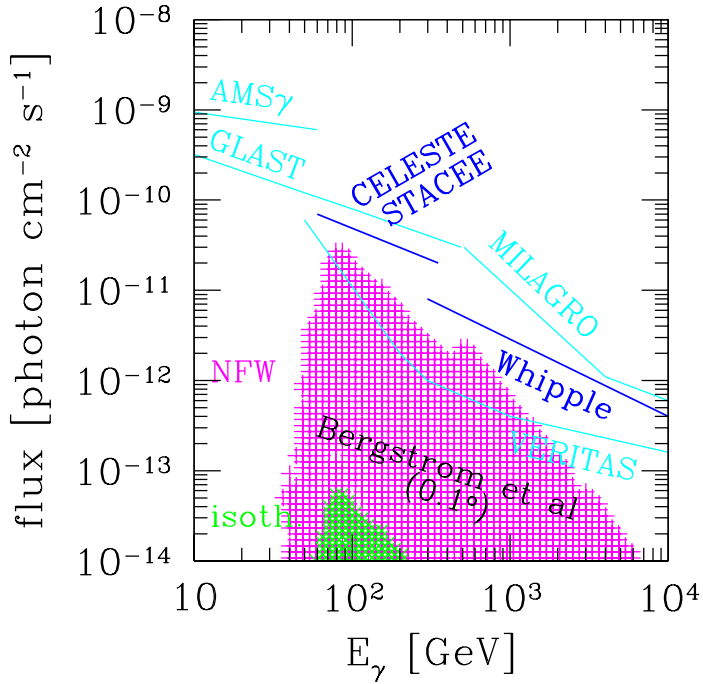}
\end{minipage}
\hfill
\begin{minipage}{0.38\textwidth}
\vskip-20pt
\caption{Expected gamma-ray flux in the gamma-ray line from neutralino annihilation in our galactic halo, coming from the direction of the Galactic Center (without the spike discussed in next Section). The photon energy $E_\gamma$ is to good approximation equal to the neutralino mass. The upper set of points is for an NFW model \protect\citep{BUB:1996}, the lower set for a cored isothermal model. The gamma-ray fluxes differ by a factor of about 500. Also shown for approximate comparison are the sensitivities of current and upcoming gamma-ray telescopes to gamma-ray point sources. (Figure from \citealp{Gondolo:2000}.)}
\end{minipage}
\end{figure}

It is therefore crucial for our purpose to know which class of profiles, with a core or with a cusp, better resembles reality. Unfortunately, constraining a given dark matter profile using the kinematics of the central region is a hard problem, because the dynamics of the central regions is usually dominated by the visible matter, and when it is not, like in low surface brightness (LSB) spiral galaxies, the central parts are so small that the angular resolution of the observations is a major concern. Discrepant data and an apparent lack of universality in LSB profiles has generated endless controversies in the astrophysical community. At any rate, it must not be forgotten that the profiles mentioned above obtained in numerical simulations include cold dark matter only, and astrophysical processes connected for instance with the gas and the formation of stars may well modify the density profile of dark matter (theoretical work in this direction is not lacking).

In the face of this situation, when making predictions for annihilation signals from neutralino dark matter in the halo, it is prudent for the moment to consider both possibilities, core and cusp. 

The considerable differences in indirect neutralino signals between a cored and a cuspy density profile are illustrated in Figure 21 for a gamma-ray signal from neutralino annihilation from the direction of the Galactic Center (without the spike contribution discussed in the next Section). There is a factor of 500 difference between assuming an NFW profile or a cored isothermal profile in the theoretical calculation of the gamma-ray flux in the seven-parameter MSSM model in \citet{BUB:1996}. Superposed on the plot are the sensitivities to gamma-ray point sources of various gamma-ray telescopes, both current and upcoming (as of 2000). The comparison with the theoretical expectations in the figure is not direct, however, because the neutralino emission may not be point like in some of the telescopes. The sensitivity curves are thus only meant to provide an approximate comparison.

It must be mentioned here that EGRET has detected gamma-ray emission from the direction of the Galactic Center \citep{Mayer-Hasselwander:1998}. However, in a reanalysis of the EGRET data by \citet{Hooper:2002}, the EGRET signal seems to originate from a source which is displaced with respect to the Galactic Center. \citet{Hooper:2002} use their reanalysis to place constraints on the gamma-ray flux from neutralino annihilations, coming to an upper limit on the gamma-ray flux above 1 GeV which is about  a factor of 2 higher than the theoretical predictions for neutralino masses $m_\chi \sim 50$ GeV, assuming an NFW profile. See \citet{Hooper:2002} for details.

\subsubsection*{Effect of halo substructure}

The same numerical simulations that predict a cuspy dark halo profile also predict the existence of numerous dark clumps in galactic halos. These clumps are a natural outcome of the hierarchal formation of structure in the Universe. Small structures form first, and larger structures, like galaxies, grow by attracting and swallowing smaller structures. This process continues to the present day. Clumps that fall into a galaxy are pulled apart by tidal interactions: the material pulled out forms tidal streams that crisscross the galactic halo. The central parts of some of the clumps may survive for a long time, and become a galactic satellite.

The overall picture of hierarchical structure formation has found confirmation in a variety of context, from observations of galaxy clusters and of merging galaxies to the halo substructure detected in our own galaxy (see the discussion on the Sagittarius stream in Section 3.1 above, for example). A numerical discrepancy should however be mentioned. A counting of visible satellites of our own galaxy gives a number of luminous satellites that is much smaller than the expected number of dark satellites. A resolution to this discrepancy may be that only a small fraction of dark satellites becomes luminous. It is not clear why this should happen, but on the other hand we do not fully understand how galaxies become luminous in the first place. 
Thus for what concerns signals from neutralino annihilation, it makes sense to examine the effect of adding substructure, i.e.\ clumps and streams, to galactic dark halos. 

\begin{figure}[!t]
\begin{minipage}{0.6\textwidth}
\includegraphics*[width=0.9\textwidth]{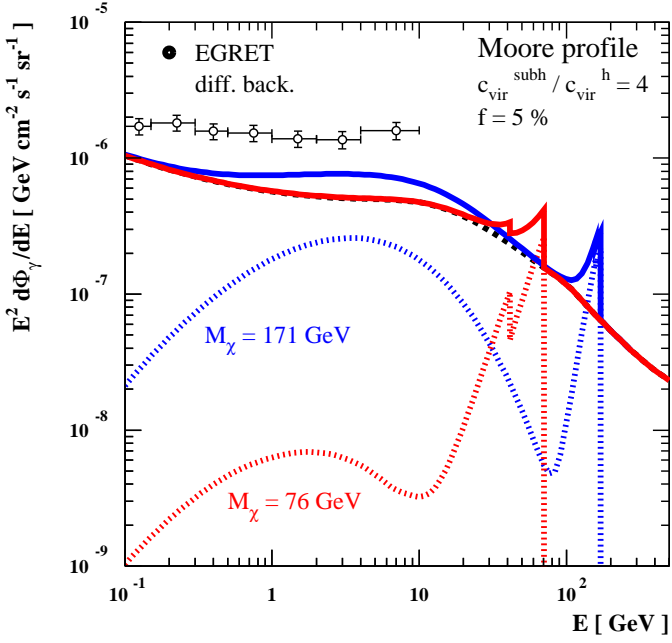}
\end{minipage}
\hfill
\begin{minipage}{0.39\textwidth}
\caption{Expected isotropic gamma-ray background from neutralino annihilations in the early Universe. Dotted lines are the signal from neutralino annihilations; solid lines are the sum of the neutralino signal and a gamma-ray background of astrophysical origin. Two neutralino models are shown, together with the current EGRET measurements of the isotropic extragalactic background. The assumptions used for the dark matter profile are indicated. (Figure from \citealp{Ullio:2002}.)}
\end{minipage}
\end{figure}

Substructure in the halo tends to increase the annihilation signals because of the dependence of the annihilation rate on the square of the dark matter density (see Eq.~(\ref{eq:gamma_ann})). The enhancement factor is linearly proportional to the density enhancement $\delta \equiv \rho'/\rho$. To understand why the dependence is linear instead of quadratic in $\delta$, consider a box of volume $V$ containing a total mass  $M$. The density in the box is $\rho=M/V$ and the annihilation rate integrated over the whole box is 
\begin{equation}
R_{\rm ann} = \Gamma_{\rm ann} V = \frac{ \sigma_{\rm ann} v }{ m^2}  \frac{M^2}{V} .
\end{equation}
 Now let all the mass be concentrated equally into $N$ small boxes, each of volume $V'$, so that each box contains a mass $M' = M/N$ and thus has a density $\rho'=M'/V'$. The density enhancement is then $ \delta = V/NV' $. The annihilation rate from the whole box is now given by a sum over the $N$ small boxes as
\begin{equation}
R_{\rm ann}' = N \Gamma'_{\rm ann} V' = N \frac{ \sigma_{\rm ann} v}{ m^2} \frac{M'^2}{V'} = \frac{\sigma_{\rm ann} v}{m^2} \frac{M^2}{V} \frac{V}{NV'} = R_{\rm ann} \delta.
\end{equation}
One power of the density $\rho$ in $\Gamma_{\rm ann}$ is compensated by a decrease in the volume where annihilations occur. Hence the signal enhancement is linear in the density increase.

\begin{figure}[!b]
\includegraphics[width=0.49\textwidth]{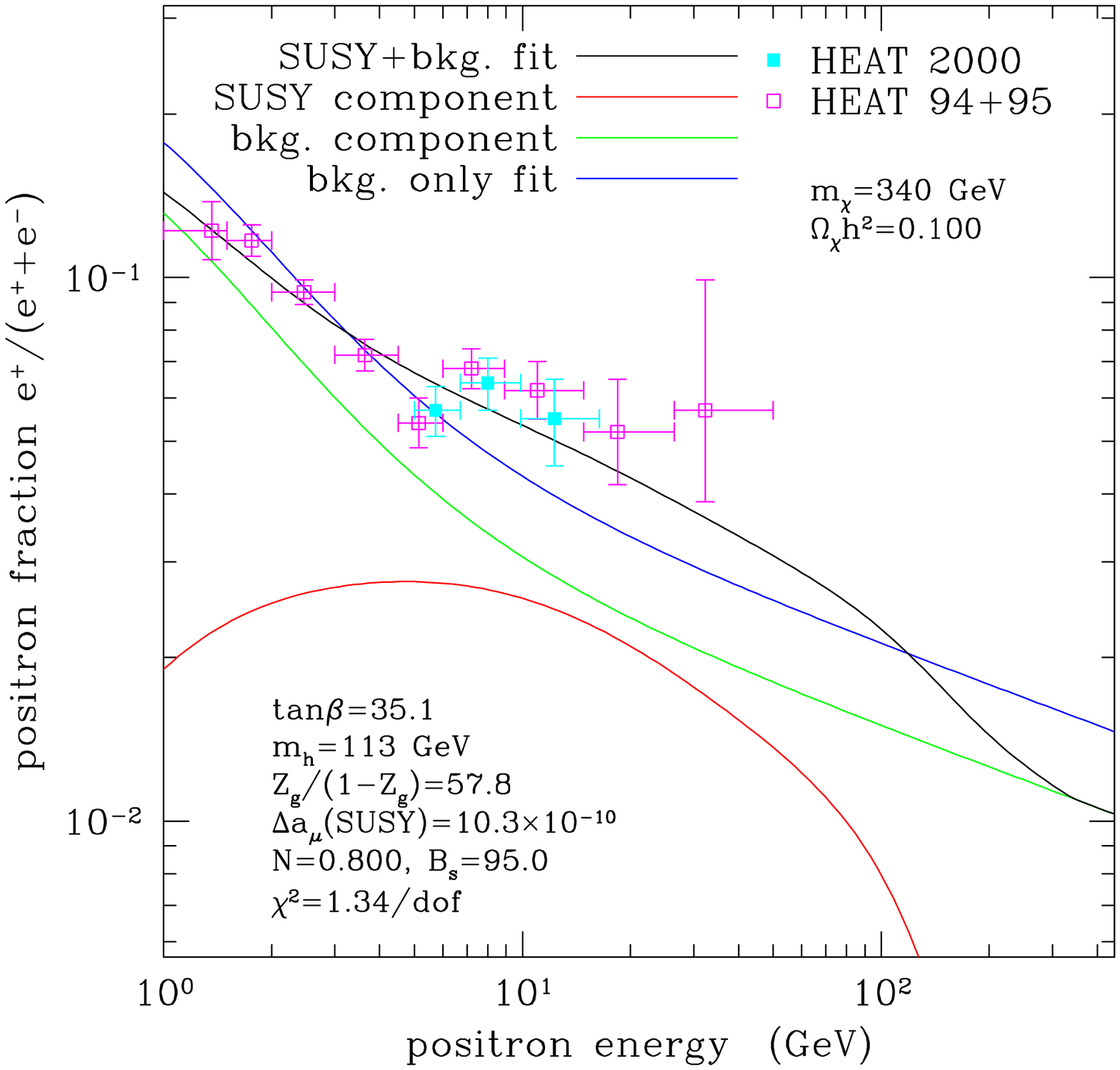}
\hfill
\includegraphics[width=0.49\textwidth]{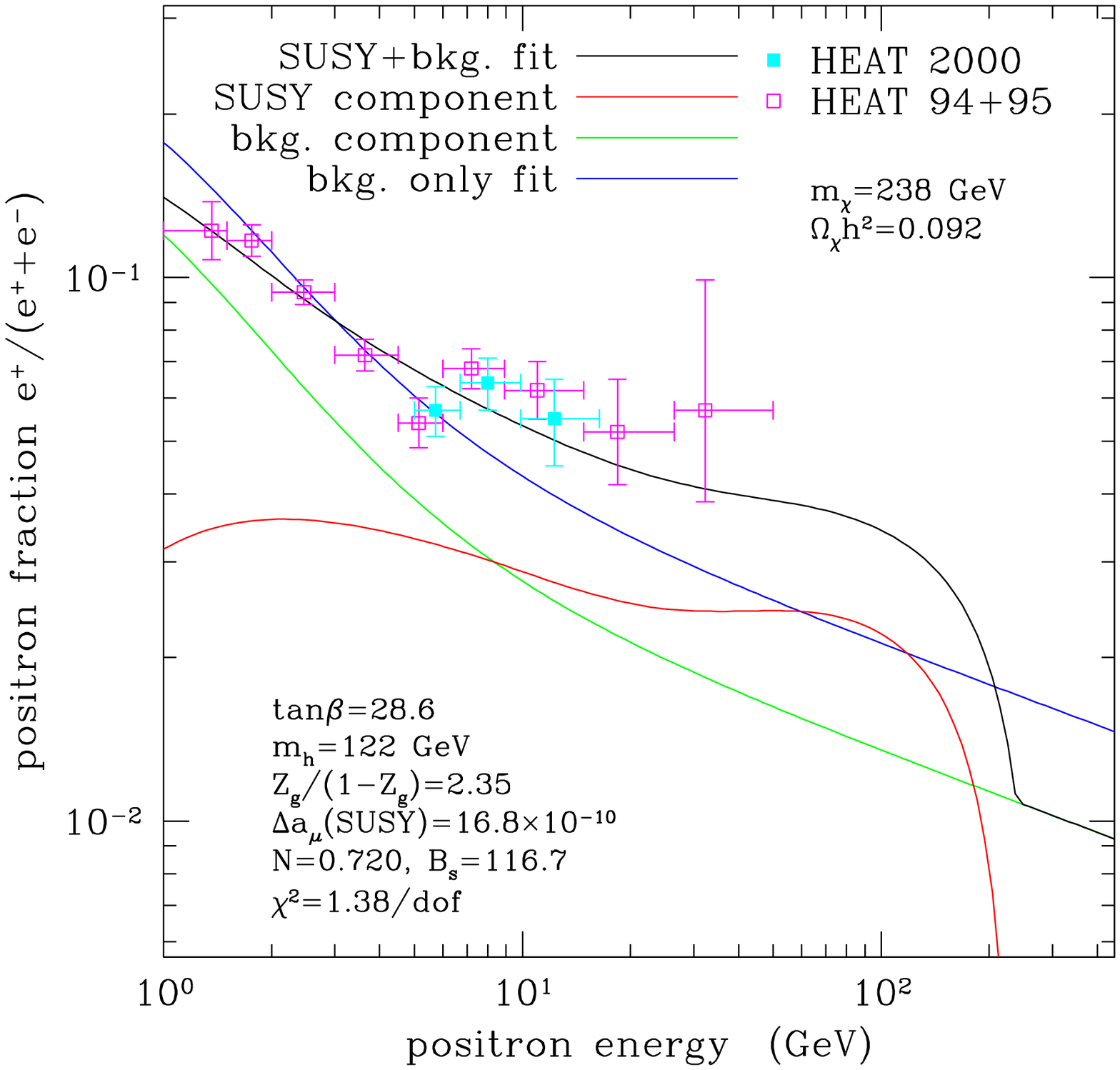}
\vspace{-20pt}
\caption{Two examples of neutralino models that provide a good fit to the excess of cosmic ray positrons observed by the HEAT collaboration. The two sets of data points (open and filled squares) are derived from two different instruments flown  in 1994-95 and 2000. The lines represent: (i) the best expectation we have from models of cosmic ray propagation in the galaxy (`bkg.\ only fit'), which underestimate the data points above $\sim$7 GeV; (ii) the effect of adding positrons from neutralino annihilations (lines `SUSY component', `SUSY+bkg.\ fit', and `bkg.\ component', the latter being the resulting background component when the data are fitted to the sum of background and neutralino contributions).
(Figures from \citealp{positrons}).}
\end{figure}

An interesting consequence of the annihilation signal enhancement has been explored by \citet{Ullio:2001} and \citet{Ullio:2002}. These authors have found that the density enhancements produced during the formation of the large scale structure in the Universe may lead to a substantial increase in the isotropic gamma-ray background from neutralino annihilations in the early Universe. Moreover, they found that this increase in the gamma-ray signal is not very sensitive to details of the galactic density profile. Expected gamma-ray spectra may in some models be close to the measured gamma-ray background, as illustrated in Figure 22. The gamma-ray spectra include both a continuum part and gamma-ray lines (two for each neutralino case: one for $\chi\chi\to\gamma\gamma$, the other for $\chi\chi\to Z\gamma$). The gamma-ray lines are asymmetrically broadened because photons emitted at earlier times have a larger redshift. The gamma-ray background due to neutralino annihilations should be searched for at high galactic latitudes, where the galactic emission is expected to be minimal. Detection of the line features depicted in Figure 22 would not require an energy resolution much better than the present one.

Another exciting application of a signal enhancement due to clumps in the galactic halo is the boost of the positron signal from neutralino annihilation up to the level of the excess of cosmic ray positrons observed by the HEAT collaboration. The HEAT collaboration flew two different detectors on balloons, and claims to have detected a ratio of positron to electron fluxes above $\sim$7 GeV that is higher than the flux expected in current state-of-the-art models of cosmic ray production and propagation in the galaxy. These models aim at explaining all correlated signals in gamma-rays, radio waves, protons, electrons, positrons, heavy nuclei, etc. that are produced by cosmic rays in our galaxy.

\begin{figure}[!t]
\centering
\includegraphics[width=0.59\textwidth]{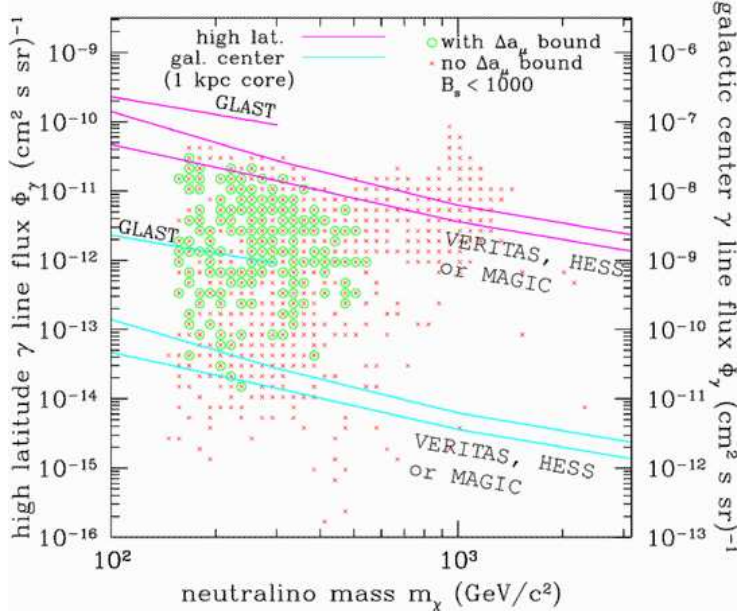}
\vspace{-5pt}
\caption{Sensitivity of upcoming gamma-ray telescopes to neutralino models that can explain the HEAT positron excess with neutralino clumps in the galactic halo. Model points are indicated by crosses; circles denote those models that in addition can also account for the measured deviation in the muon magnetic moment. The upper set of sensitivity curves corresponds to the high latitude gamma-ray line flux (scale on the left); the lower set of curves to the direction toward the galactic center (scale on the right; no steep spike around the central black hole is assumed). (Figure from \citealp{positrons}.)}
\vspace{-15pt}
\end{figure}

One possible explanation is that the positron excess is due to the positrons generated in neutralino annihilation in the galactic halo. If the neutralinos are produced thermally in the early universe, which is the most common assumption, the annihilation cross section $\sigma_{\rm ann}$ is forced to be small by the requirement that the neutralino relic density is large enough for neutralinos to be the dark matter. Using the average value of the local dark matter density, of the order of 0.3 GeV/cm$^2$, then leads to a positron signal which is more than an order of magnitude smaller than the excess measured by HEAT. Increasing the annihilation cross section $\sigma_{\rm ann}$ does not make the signal higher, because the density $\rho$ decreases inversely with $\sigma_{\rm ann}$, and hence the annihilation signal, being proportional to $\sigma_{\rm ann} \rho^2$, decreases. 
\citet{Kane:2001} suggested that the neutralinos may not have been produced thermally in the early Universe, and were thus able to decouple the annihilation rate in the halo from the constraint coming out of the relic density requirement. Alternatively, \citet{positrons} have suggested that substructure in the galactic halo may provide the necessary boost factor to the positron signal. 

Enhancing the positron signal through clumps also enhances other annihilation signals, such as antiprotons and gamma-rays. \citet{positrons} have performed a detailed analysis of these enhancements, and have concluded that it is possible to explain the HEAT positron excess with boost factors as small as 30, but typically higher, without obtaining too many antiprotons or gamma-rays. Figure 23 shows two examples of neutralino models that provide a good fit to the positron excess. On the left, the neutralino has mass $m_\chi = 340 $ GeV and is an almost pure gaugino (gaugino fraction $Z_g  = 0.98$); the boost factor is 95 and other parameters are listed in the figure. On the right, the neutralino is mixed  (gaugino fraction $Z_g = 0.70$) with a mass of 238 GeV; the boost factor is 116.7. The $\chi^2$ per degree of freedom is quite good for both fits, 1.34 and 1.38 respectively.

The ultimate test of the explanation of the positron excess by means of neutralino clumps will be the detection of a signal in gamma-rays. Gamma-ray production would in fact be enhanced by the same mechanism that would enhance positron production. \citet{positrons} have found that almost all neutralino models that can explain the positron excess are within the sensitivity reach of upcoming gamma-ray telescopes (see Figure 24). The realistic possibility of confirming (or disproving) the neutralino origin of the positron excess is fascinating.

\subsubsection{Signals from neutralino annihilation at the Galactic Center}

The last indirect neutralino signals we consider are neutrinos, gamma-rays and radio waves from a possible dark matter concentration around the black hole at the Galactic Center. 

Evidence for the presence of a black hole at the center of our galaxy comes from studies of the motion of stars in orbit around the center. The speeds of these stars decrease from the center as the inverse square root of the radius, which is the primary indication for the existence of a point mass at the center. The mass of the central object is measured to be $\sim 4 \times 10^6$ solar masses, which are contained within a sphere of less than $\sim 0.05$ pc \citep{Eckart:1997,Ghez:1998,Ghez:2003}. No stellar or gas system can be so dense, indicating that the central object is most probably a black hole. The position of the black hole happens to coincide with the position of a strong radio source called Sagittarius A$^*$, which is thus identified with the central black hole.

The radio emission from Sgr A$^*$ is easily explained by thermal emission of hot matter falling into the black hole. However, contrary to many of the similar black holes observed at the center of external galaxies, our galactic black hole does not emit intensely in the X-ray band, and it is controversial if it emits gamma-rays.              Models for such `quiet' black holes do exist, however, such as those involving advection-dominated accretion flows (ADAFs). 

Further evidence for a black hole at the center of the Milky Way comes from the 2001 observation of a X-ray and infrared flares from the Galactic Center \citep{Baganoff:2001,Porquet:2003,Genzel:2003b}. The flare time scale and intensity can nicely be explained if the flare is produced near a black hole \citep[see, e.g.,][]{Aschenbach:2004}.

\begin{figure}[!t]
\centering
\includegraphics[width=0.7\textwidth]{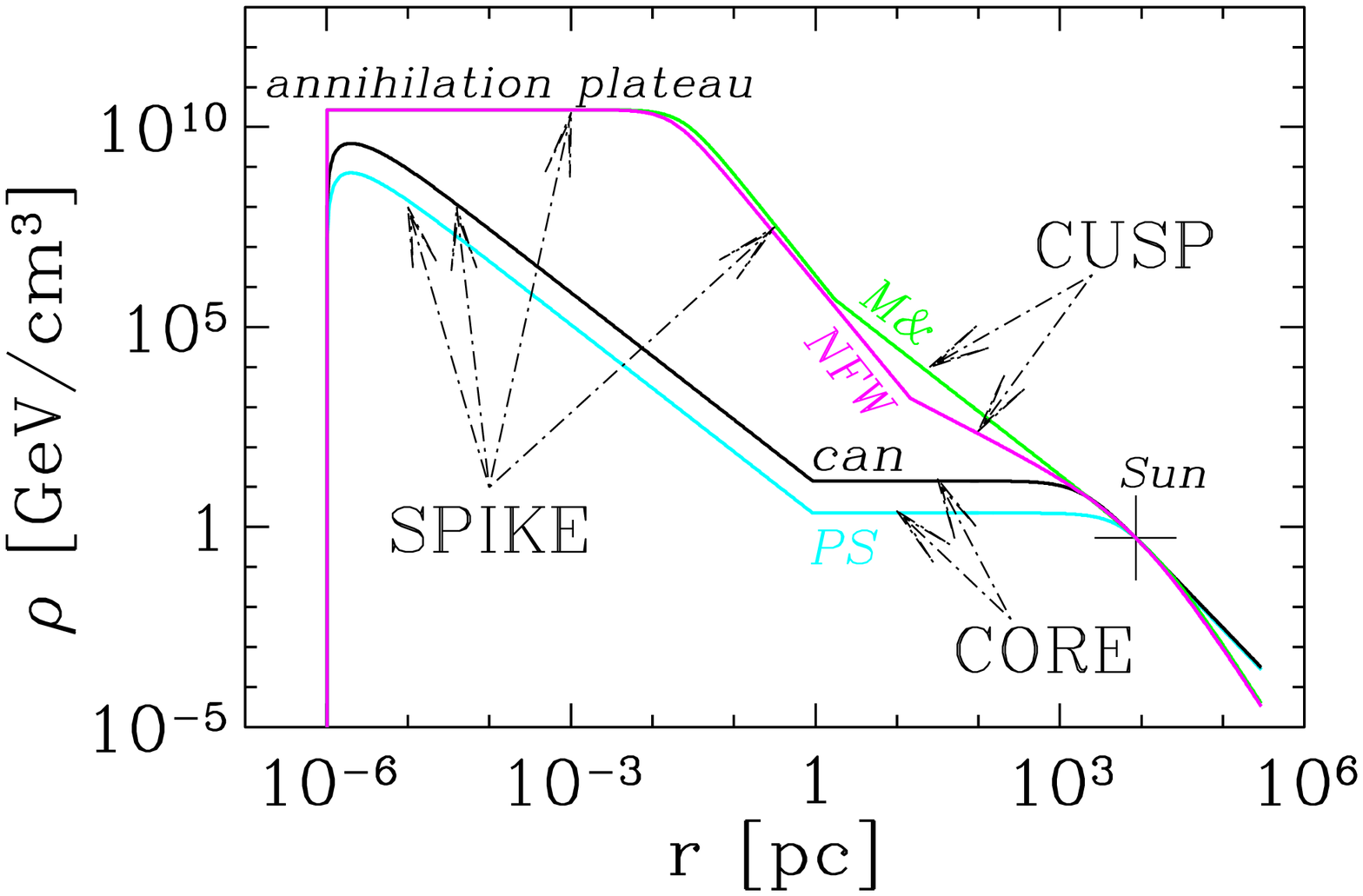}
\caption{Density profiles of spikes that form adiabatically around the black hole at the center of our Galaxy. The position of the Sun is indicated by a cross. Four models for the halo profile are shown: two with cores (`PS' by \citet{Persic:1997} and `can' by \citet{Bahcall:1980}) and two with cusps (`NFW' by \citet{Navarro:1996} and `M\&' by \citet{Moore:1998}). The spikes form within the radius of influence of the black hole, $r_{\rm infl} \sim 1$ pc. In the `annihilation plateau' neutralino annihilations have been so rapid as to deplete the number of neutralinos. (Figure from \citealp{Gondolo:Snowmass}.)}
\end{figure}

Dark matter may be driven near the black hole gravitationally, and may form a dense concentration around it \citep{Gondolo:1999}. We will call this concentration a spike, so as to distinguish it from the dark matter cusps of Section 3.2.2.  The formation of a spike is gravitational phenomenon similar to but less efficient than accretion of matter, in that the latter involves dissipation of energy and angular momentum and can thus produce concentrations which are smaller and denser.

How strong is the dark matter concentration around Sgr A$^*$, or around a generic black hole? This is still a matter of investigation. The simplest case is that of adiabatic compression, and was analyzed by \citet{Gondolo:1999}. 

Adiabatic compression of an initial dark matter distribution produces two kinds of spikes. If the initial distribution before the black hole forms is thermal, the spike is shallow, with density profile $\rho \propto r^{-3/2}$. If the initial phase-space distribution is  a power-law in energy, the spike is steep, i.e.\ $\rho \propto r^{-\gamma}$ with $\gamma>2$. The physical reason for the higher concentration in the second case is the presence of many dark matter orbits with low speed, which are more easily driven into bound orbits when the black hole forms. 

These two kinds of spike are illustrated in Figure 25 for the same four halo models discussed in Section 3.2.2. Models with a core, like those by \citet{Bahcall:1980} and \citet{Persic:1997}, give rise to a shallow spike around the central black hole, while models with a cusp, like those of \citet{Navarro:1996} and \citet{Moore:1998}, produce steep spikes. In the very inner regions, the density may become so high that neutralino-neutralino annihilations may have had the time to deplete the number of neutralinos and an `annihilation plateau' is formed. The typical radius of a spike around a black hole is determined by the radius of influence of the black hole, $r_{\rm infl} \sim G M/\sigma_v^2 $, which is the radius at which the gravitational potential energy becomes comparable to the typical kinetic energy in the dark matter gas ($M$ is the black hole mass and $\sigma_v$ is the gas velocity dispersion). For the black hole at the Galactic Center, $r_{\rm infl} \sim 1$ pc.

Neutralino annihilation is enhanced in the spike, because of the dependence of the annihilation rate on square of the density. The Galactic Center then becomes a source of neutrinos, gamma-rays, radio waves, etc.\ from neutralino annihilation (Figure 26). The intensity of the emission depends on the steepness of the spike. If the spike is shallow, neutralino annihilation is generally undetectable. On the contrary, a steep spike at the Galactic Center produces interesting signals. 

\begin{figure}[!b]
\begin{minipage}{0.5\textwidth}
\includegraphics[width=\textwidth]{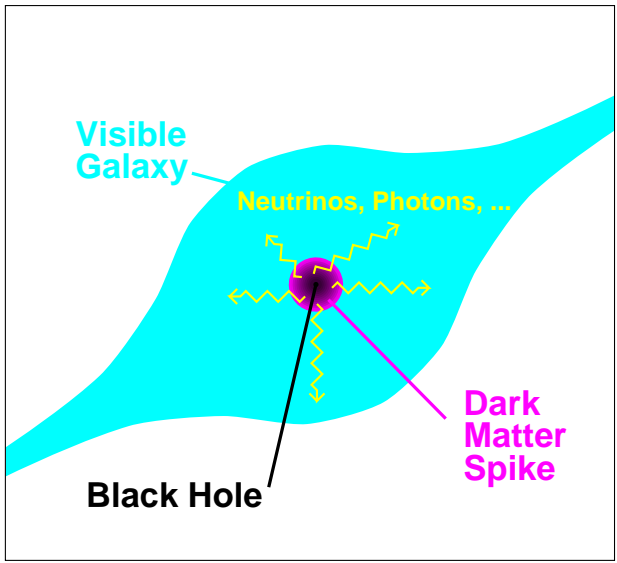}
\end{minipage}
\hfill
\begin{minipage}{0.4\textwidth}
\caption{Artistic conception of emission from a dark matter spike around a black hole at the galactic center. Neutralino annihilation in the spike produces intense fluxes of neutrinos, gamma-rays, radio waves, etc. Some of these signals may be detectable.}
\end{minipage}
\end{figure}

For example, if, disregarding Hooper \& Dingus's reanalysis mentioned above, we attribute the EGRET gamma-ray emission from the Galactic Center to neutralino annihilation in a spike born out of an NFW profile, we obtain a high-energy neutrino flux that is either excluded or mostly detectable in a km$^3$ neutrino telescope (Figure 27). The flux of neutrino-induced muons above 25 GeV would be detectable over the atmospheric neutrino background for neutralino masses between $\sim$100 GeV and $\sim$2 TeV, while heavier neutralinos would already be excluded from the current limit on the neutrino emission from the Galactic Center \citep{Habig:2001}.

\begin{figure}[!b]
\centering
\includegraphics[width=0.6\textwidth]{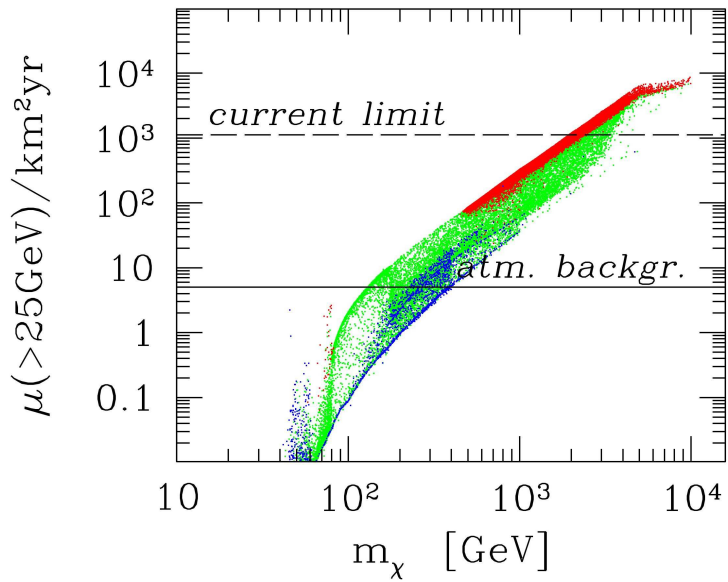}
\caption{Predicted neutrino signal from neutralino dark matter annihilation in a steep adiabatic spike at the Galactic Center expressed as number of neutrino-induced muons per km$^2$-yr in a neutrino telescope. The spike corresponds to an NFW profile. Each dot in the figure corresponds to a point in a seven-parameter weak-scale MSSM, and is normalized so that the gamma-ray flux from the spike coincides with the gamma-ray signal from the Galactic Center observed by EGRET. Models with heavy neutralinos are excluded by the current limit on the neutrino emission from the Galactic Center; models with neutralinos as light as $\sim$100 GeV could be detected above the atmospheric neutrino background. (Figure from \citealp{Gondolo:NSF02}.)}
\end{figure}

As a second and more dramatic example \citep{Gondolo:2000b},  electrons and positrons from neutralino annihilation would emit synchrotron radiation as they spiral in the magnetic field that plausibly exists around the central black hole. While this synchrotron radiation is innocuous for a shallow adiabatic spike, it may exceed the observed radio emission by several orders of magnitude if the spike is steep. The radio synchrotron emission at 408 MHz is shown in Figure 28 for
an adiabatic spike born out of an NFW profile, under two assumptions for the radial dependence of the magnetic field (a constant field of 1 mG and a field in equipartition with the infalling gas). In both cases, all dark matter neutralinos in the seven-parameter MSSM models considered are strongly excluded.

\begin{figure}[t]
\begin{minipage}{0.39\textwidth}
\caption{Synchrotron emission at 408 MHz expected from neutralino dark matter annihilation in a steep adiabatic spike at the Galactic Center. The spike corresponds to an NFW profile, and the synchrotron radiation is emitted by electrons and positrons produced in neutralino annihilation. Upper panel: constant magnetic field; lower panel: equipartition magnetic field. All dark matter neutralinos in the seven-parameter weak-scale MSSM  considered are excluded by several orders of magnitude. (Figure from \citealp{Gondolo:2000b}.)}
\end{minipage}
\hfill
\begin{minipage}{0.6\textwidth}
\includegraphics[width=\textwidth]{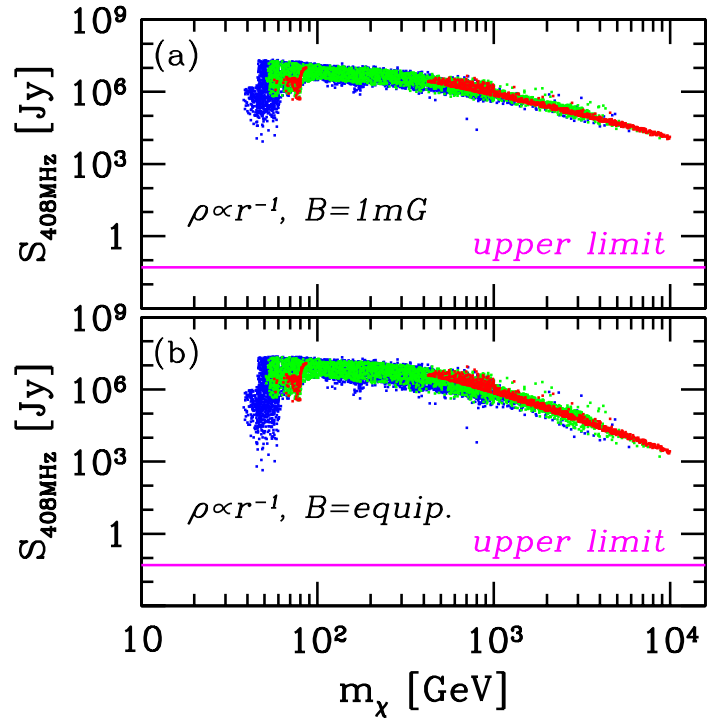}
\end{minipage}
\end{figure}

The examples above have assumed that the spike formed adiabatically and maintained its shape till the present time. This may not be the case. 

For example, if the central black hole formed through the merging of two black holes of similar mass, \citet{Merritt:2002} have shown that the spike  would become shallow at the end of the merging, because
dark matter particles would be kicked out of the black hole region via a gravitational sling-shot effect. The final shallow spike would not give dramatic signals from neutralino annihilation at the Galactic Center. In a realistic scenario of this type, the two merging black holes of similar mass would be accompanied by their host galaxies whose mass would also be similar. The merging would then constitute what is known as a major merging. A major merging is capable of destroying the Galactic disk, and so must not have occurred after the disk formed about $10^{10}$ years ago. Thus, for this scenario to work, black holes of millions of solar masses should already have been in place at very early times. Is this possible? While hard to explain theoretically, supermassive black holes have indeed been observed in very distant quasars, at redshift $z>6$, so the scenario may be plausible.

In these considerations, it must be kept in mind that there is a stellar spike around the black hole at the Galactic Center. The steepness of this stellar spike is however not very well know. With large uncertainties, \citet{Alexander:2000} estimate the slope of the stellar spike to be $\gamma_{\rm stars} \sim 1.3$--1.4. This means that the current stellar spike is probably shallow. We may think that the stellar spike is our best proxy for the dark matter spike. If so, also the dark matter spike would also be shallow, and thus inconsequential for neutralino signals. However, the dark matter and stellar spikes follow very different evolution histories, because contrary to the dark matter, binary collisions of stars and coalescence of two stars into one at collisions effectively relax the stellar system to a shallower spike. 

The final word has not yet been said regarding the distribution of dark matter at the Galactic Center, or around black holes and other compact objects in general. This is one of the exciting points of contact between the study of dark matter and the study of the formation and evolution of galactic nuclei.

\section{Conclusions}

Current cosmological data imply the existence of non-baryonic dark matter. 
We have discussed some of the most popular candidates and shown that none of the candidates known to exist, i.e.\ the active neutrinos, can be non-baryonic cold dark matter. Hence to explain the nature of cold dark matter we need to invoke hypothetical particles that have not been detected yet. Some of these hypothetical particles have been suggested for reasons different than the dark matter problem (such as sterile neutrinos, neutralinos, and axions), some others have been proposed mainly as a solution to the cold dark matter problem (e.g., self-interacting dark matter, WIMPZILLAs, etc.). Although most studies focus on the first category of candidates, especially neutralinos and axions, we should keep an open mind. 

To illustrate how we can find out if dark matter is made of elementary particles, 
we have used neutralino dark matter as our guinea pig to survey several methods  to search for non-baryonic dark matter. These methods range from a direct detection of dark matter particles in the laboratory to indirect observation of their annihilation products produced in the core of the Sun or of the Earth and in galactic halos, including our own. Direct searches may have found a signal from WIMPs (the annual modulation), but this claim is highly controversial at the moment. Future direct searches have great promise, and might even be able to explore the local velocity distribution of WIMPs. These searches are complemented by indirect searches for high-energy neutrinos from the core of the Sun or of the Earth. Indirect searches using gamma-rays and cosmic rays from annihilations in the galactic halo are subject to uncertainties related to the detailed structure of the dark matter halo. Even more so are predictions for dark matter signals from the Galactic Center. Despite this, some anomalies in cosmic ray fluxes, namely a positron excess, may be explained by neutralino annihilation, and future gamma-ray observations may discover a gamma-ray line from neutralino annihilation in our galactic halo.

All of the examples we have presented are without doubt simple, elegant, and compelling explanations for the nature of non-baryonic dark matter. As we ponder on which one of them is realized in Nature, we must remember the words of astrophysicist Thomas Gold
(as quoted by Rocky Kolb): ``For every complex natural phenomenon there is a simple, elegant, compelling, wrong explanation."

\end{document}